\documentclass[pre,twocolumn,amsmath,amssymb,notitlepage]{revtex4}

\usepackage{dcolumn}
\usepackage{lipsum}
\usepackage{graphicx}
\usepackage{hyperref}
\usepackage{float}
\usepackage{bm}
\usepackage[T1]{fontenc}
\usepackage{color}
\usepackage{url}
\usepackage[bf]{subfigure}
\usepackage{rotating}

\begin{document}

\title{Concentric Network Symmetry}

\author{Filipi N. Silva$^1$}
\author{Cesar H. Comin$^1$}
\email{chcomin@gmail.com}
\author{Thomas K. DM. Peron$^1$}
\author{Francisco A. Rodrigues$^2$}
\author{Cheng Ye$^3$}
\author{Richard C. Wilson$^3$}
\author{Edwin Hancock$^3$}
\author{Luciano da F. Costa$^1$}
\affiliation{$^1$Instituto de F\'{\i}sica de S\~{a}o Carlos, Universidade de S\~{a}o Paulo, S\~{a}o Carlos, S\~ao Paulo, Brazil\\
$^2$Departamento de Matem\'{a}tica Aplicada e Estat\'{i}stica, Instituto de Ci\^{e}ncias Matem\'{a}ticas e de Computa\c{c}\~{a}o, Universidade de S\~{a}o Paulo, Caixa Postal 668,13560-970 S\~{a}o Carlos, S\~ao Paulo, Brazil\\
$^3$Department of Computer Science, University of York, York, YO10 5GH, United Kingdom}


\begin{abstract}
Quantification of symmetries in complex networks is typically done globally in terms of automorphisms. Extending previous methods to locally assess the symmetry of nodes is not straightforward. Here we present a new framework to quantify the symmetries around nodes, which we call connectivity patterns. We develop two topological transformations that allow a concise characterization of the different types of symmetry appearing on networks and apply these concepts to six network models, namely the Erd\H{o}s-R\'enyi, Barab\'asi-Albert, random geometric graph, Waxman, Voronoi and rewired Voronoi. Real-world networks, namely the scientific areas of Wikipedia, the world-wide airport network and the street networks of Oldenburg and San Joaquin, are also analyzed in terms of the proposed symmetry measurements. Several interesting results emerge from this analysis, including the high symmetry exhibited by the Erd\H{o}s-R\'enyi model. Additionally, we found that the proposed measurements present low correlation with other traditional metrics, such as node degree and betweenness centrality. Principal component analysis is used to combine all the results, revealing that the concepts presented here have substantial potential to also characterize networks at a global scale.
\end{abstract}

\maketitle

\section{Introduction}

The study of graph symmetry has its roots in algebraic graph theory~\cite{godsil2001algebraic,scapellato2003topics,biggs1993algebraic,beineke2004topics}. In this context, symmetry is defined in terms of \textit{automorphisms} and the \textit{orbits} of the nodes in a given graph. More specifically, an automorphism of a graph $G$ is a permutation of node labels which preserves the label adjacency relation of the original graph. The orbit of a given node is the set of remaining nodes that share a label over the permutations in a automorphism group ~\cite{godsil2001algebraic,scapellato2003topics,biggs1993algebraic,beineke2004topics}. A network is assumed to be symmetric if it presents at least one node label permutation such that the label adjacency remains unchanged. Otherwise, the network is considered to be asymmetric~\cite{godsil2001algebraic,scapellato2003topics,biggs1993algebraic,beineke2004topics,xiao2008emergence}.

Despite the great amount of work devoted to the structural and dynamical characterization of complex networks, the study of symmetry has received less attention, including works such as~\cite{holme2006detecting,xiao2008emergence,macarthur2008symmetry}. In algebraic graph theory, the analysis is focused on simple graphs. More recently, measurements based on automorphisms and orbits have been employed to analyse the structure of complex networks. For instance, Xiao et al.~\cite{xiao2008emergence} used the size of automorphisms groups~\cite{macarthur2008symmetry} to quantify the symmetry of real-world complex networks, showing that the emergence of symmetry in such networks is a consequence of similar patterns of connectivity. Furthermore, in~\cite{xiao2008emergence} the authors proposed a variation of the preferential-attachment mechanism in which new nodes are connected with probability proportional to the degree of nodes together as well as their local symmetry. Interestingly, the model was capable of reproducing the symmetry properties observed in real-world networks, reinforcing the hypothesis that similar linkage properties are crucial for the emergence of overall symmetry~\cite{xiao2008emergence}. 

While symmetry measurements based on automorphisms typically apply to the whole graph, it is also possible to develop approaches aimed at quantifying symmetry locally, for instance in terms of a given reference node. One of the first studies of local symmetry in complex networks was reported by Holme~\cite{holme2006detecting}, where the so-called \textit{degree-symmetry} of a node was proposed. Although this allowed the local symmetry to be quantified with respect to a reference node, occasional negative values of the proposed measurement can be obtained. In addition, it does not posses clear bounds, making the comparison of the symmetry between patterns with different number of nodes less direct. 

Quantum walks can also be used to detect both exact and partial local symmetry in networks. It has long been known that quantum walks hit exponentially faster in networks that possess symmetry \cite{kempe_quantum_2003,kempe_quantum_2003_2,krovi_hitting_2006}. Emms et al \cite{emms_graph_2009} have shown how to characterize quantum walks in terms of a measure of commute time (the sum of expected value of outward and return hitting times between a pair of nodes). When the walk is represented by an embedding into a vector
space where the Euclidean distance between nodes is the commute time, they show that in certain subspaces the positions of the nodes are degenerate, and that  these subspaces correspond to symmetries of the network~\cite{emms_graph_2009}. The reason for the degeneracy are interference effects which lead to cancellation of the wave-function due to symmetries in the network. Recently, Rossi et al \cite{rossi_characterizing_2013} have  exploited the link between destructive inference and symmetry to develop a method for partial symmetry detection in networks. The method relies on comparing pairs of walks using the quantum Jensen-Shannon divergence to measure their interference. Compared to the  work reported herein, the work of Rossi et al is computationally more expensive since it requires the simulation of multiple quantum walks.

There are several reasons why to quantity the topological symmetry around specific nodes of a network. To begin with, the intrinsic way in which several networks, synthetic or real, are constructed, tends to imply a characteristic local symmetry. For instance, patterns in city networks should reflect the grid-like organization typically adopted in urban planning. On the other hand, the hubs in scale free networks could be expected to disrupt local symmetry. Interestingly, such types of symmetries specific to different networks are also expected to impact the dynamics unfolding in the respective structure. For instance, highly symmetric patterns could be expected to promote a more uniform traffic flow in a city.

\begin{figure}[!tpb]
  \begin{center}
  \includegraphics[width=0.9\linewidth]{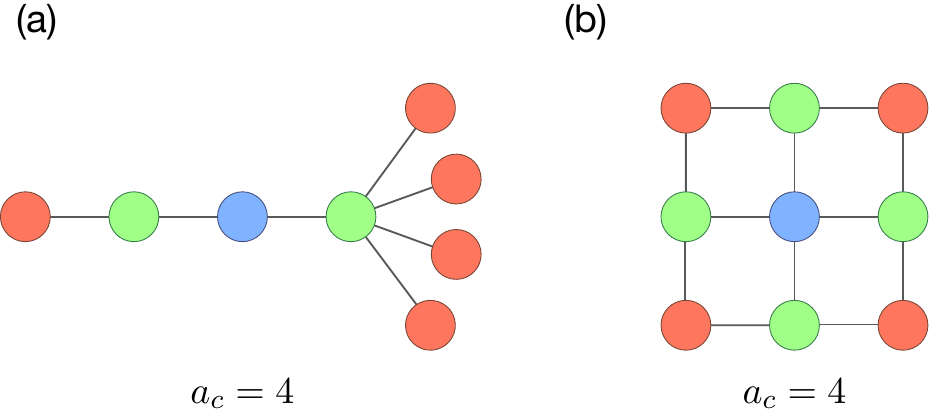} 
  \caption{(Color online) Illustration of the dual meaning of accessibility. A self-avoiding random walk dynamics is used. The graph in (a) has 5 reachable nodes but an effective number of accessed nodes equal to $a_c=4$. The graph in (b) has exactly 4 reachable nodes and since they are all equally accessed, $a_c=4$.}
  ~\label{f:motiv_norm}
  \end{center}
\end{figure}

Here, as in~\cite{holme2006detecting}, we study local symmetry across the concentric levels of a node, where a concentric level is defined as the set of nodes having the same shortest path length to the reference node. However, our definition of symmetry is based on the accessibility of a node~\cite{travenccolo2008accessibility,viana2012effective}, which is also a concentric measurement. The accessibility $a_c$ of a node for a given concentric level is calculated from the transition probabilities of the node to its neighbors at that level. These transition probabilities are defined by the dynamics imposed on the network, such as self-avoiding random walks or epidemic spreading. The accessibility is calculated by taking the exponential of the entropy of the transition probabilities \cite{viana2012effective}. Therefore, the accessibility attains its maximum value when all transition probabilities to a given concentric level are equal. In this case, the accessibility, or effective degree, of the reference node becomes equal to the number of nodes in the concentric level, indicating that the nodes can be optimally accessed. This means that the accessibility is influenced by the topological symmetry around a node, as well as the number of nodes at the given concentric level. In Fig.~\ref{f:motiv_norm} we provide an illustration of this property. For the graph shown in Fig.~\ref{f:motiv_norm}(a), a self-avoiding agent departing from the reference node (colored in blue) can reach 5 nodes in two steps, but since it is more probable that it will reach the node on the left, the reference node can \emph{effectively} access \emph{less} than 5 nodes. In this case the accessibility is $a_c=4$ (please refer to \cite{viana2012effective} for the mathematical definition of accessibility). The graph shown in Fig.~\ref{f:motiv_norm}(b) has exactly 4 reachable nodes, and the neighborhood of the reference node is perfectly symmetric. In this case we also have $a_c=4$.

The symmetry measurement presented here is, at the core, a normalized version of the accessibility, where the normalization factor is the number of reachable nodes. We combine this normalized version of accessibility with a variation of the self-avoiding random walk dynamics, referred to as the concentric random walk, to evaluate the symmetry of the node with respect to its successive concentric levels. This specific type of random walk is adopted in order to avoid transitions of moving agents to previous concentric levels, which would otherwise impair the exploration of the remainder of the network.  For the same reason, the links within a same concentric level need to be treated so that the random walk does not loose steps inside each level.  This is done by removing or merging the links inside each level, which gives rise to two respective symmetry indices. The first is the \emph{backbone symmetry}, in which the links inside a concentric level are eliminated (no transition between them). The second is the \emph{merged symmetry}, in which these links assume cost zero in the transition. As such, these two indices provide complementary quantification of the local symmetry around a given node. 

\begin{figure*}[!tpb]
  \begin{center}
  \includegraphics[width=1.0\linewidth]{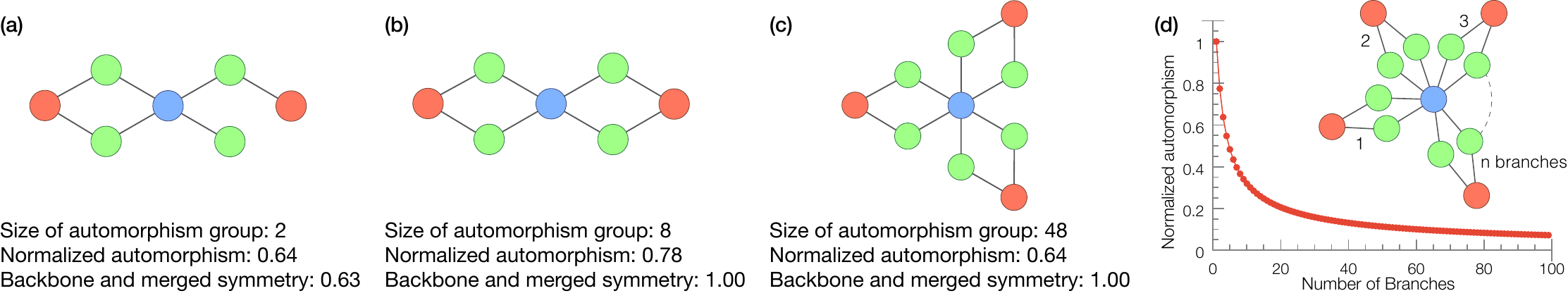} 
  \caption{(Color online) Comparison between the automorphism group size and the backbone and merged symmetry measurements. When the initial graph (a) is transformed into a more symmetric one (b), all symmetry measurements increase. In (c), some nodes are added, forming a new branch to the pattern. Although the symmetry of the pattern is maintained, the size of the automorphism group increases while the normalized automorphism decreases. The symmetry measurement proposed in the current work remains maximal. When adding even more branches, as shown in (d), the normalized automorphism tends to zero.}
  ~\label{f:motiv_mea}
  \end{center}
\end{figure*}

As mentioned above, the size of the automorphism group has been traditionally associated to the symmetry of a graph~\cite{biggs1993algebraic,xiao2008emergence}. But since the measurement has a loose upper bound, given by $N!$, defining a proper normalization to allow the comparison of symmetry levels between two given graphs is usually problematic. In Fig.~\ref{f:motiv_mea} we show a comparison between the size of the automorphism group, its normalized version and the symmetry measurements defined in the current article.   The quantification of local symmetry with respect to a reference node (colored in blue) is addressed in terms of the \emph{colored automorphism problem}, by assingning a same color to all nodes in a concentric level. The normalized automorphism measurement can then be calculated by dividing the size of the automorphism group by $N_0!N_1!N_2! \dotsm N_h!$, where $N_r$ is the number of nodes in the $r$-th concentric level, and taking the result to the power of $1/N$.  Please refer to Section \ref{s:def} for the definition of the backbone and merged symmetry measurements shown in the figure. The graph shown in Fig.~\ref{f:motiv_mea}(a) is not symmetric, implying low values for all the three respective measurements. The graph in Fig.~\ref{f:motiv_mea}(b) is seemingly more symmetric than that shown in Fig.~\ref{f:motiv_mea}(a). This is reflected in its respective measurements, since they all show larger values than for the previous cases. In this study, we consider that the graph presented in Fig.~\ref{f:motiv_mea}(b) has the highest possible symmetry for this coloring. Our measurement achieves the maximum value for this graph, while the normalized automorphism measurement presents a low value compared to its maximum. Fig.~\ref{f:motiv_mea}(c) shows another symmetric graph having additional nodes in the concentric levels, which causes the normalized automorphism measurement to decrease in value, while our measurement remains maximal. Adding more nodes to the pattern while preserving the symmetry, as shown in Fig.~\ref{f:motiv_mea}(d), decreases the normalized automorphism, whereas the backbone and merged symmetries remain maximal. In summary, the maximum size of the automorphism group is proportional to the factorial of the number of nodes at each concentric level of the pattern. Therefore, developing a proper normalization to apply this measurement in a sparse network is a difficult task.

Complex networks theory usually deals with graphs whose topology is more \emph{irregular} than what would be expected in a uniformly random model such as Erd\H{o}s-R\'enyi~\cite{costa2009seeking}. Formally, a graph is said to be regular~\cite{biggs1993algebraic} if all its nodes have the same degree, implying some level of uniformity and simplicity. However, it is known that the degree alone is not sufficient to completely characterize a graph. Thus, the term \emph{complex} in complex networks actually refers to intricate distributions not only of degree, but also of other measurements, as discussed in~\cite{costa2009seeking}. In this context, the two measurements proposed here can be understood not only as a quantification of node-centered symmetry, but also as a more robust evaluation of its regularity, and therefore \emph{simplicity}, of the network~\cite{costa2009seeking}. In other words, complex networks could be characterized and even defined as graphs possessing several nodes presenting high values of backbone and merged symmetry measurements.

This paper is organized as follows: Section II presents our modification of the accessibility measurement, which is used to define the backbone and merged symmetry measurements. Section III is devoted to the characterization of symmetry patterns emerging in random network models. The same analysis is performed for real-world networks in Section IV. In section V we perform a multivariate statistical analysis method to summarize all obtained results. Finally, conclusions are presented in Section VI.

\section{The Backbone and Merged Symmetries}
\label{s:def}

Two symmetry measurements are defined here for each concentric level $\Gamma_i^{(h)}$ of a given reference node $i$, where the concentric level is defined as the set of nodes that are at topological distance $h$ from $i$~\cite{Costa:2008p14}. For example, the first concentric level is the set $\Gamma_i^{(1)}$ of neighbors of node $i$. The symmetries are calculated in terms of the transition probabilities of a \emph{concentric random walk}. In this type of dynamics, the agent cannot return to the previous concentric level, which would be detrimental to reaching further levels.  Therefore, the edges between nodes belonging to a same concentric level should be treated in special manner. There are two ways to deal with these intra-level edges: (i) to remove then; and (ii) to join the nodes which they interconnect. As explained in more detail in the following, this gives rise to two intermediate representations of the subgraph in question, and the concentric random walk is henceforth specified with respect to these two derived structures.

\begin{figure*}[!tpb]
  \begin{center}
  \includegraphics[width=0.75\linewidth]{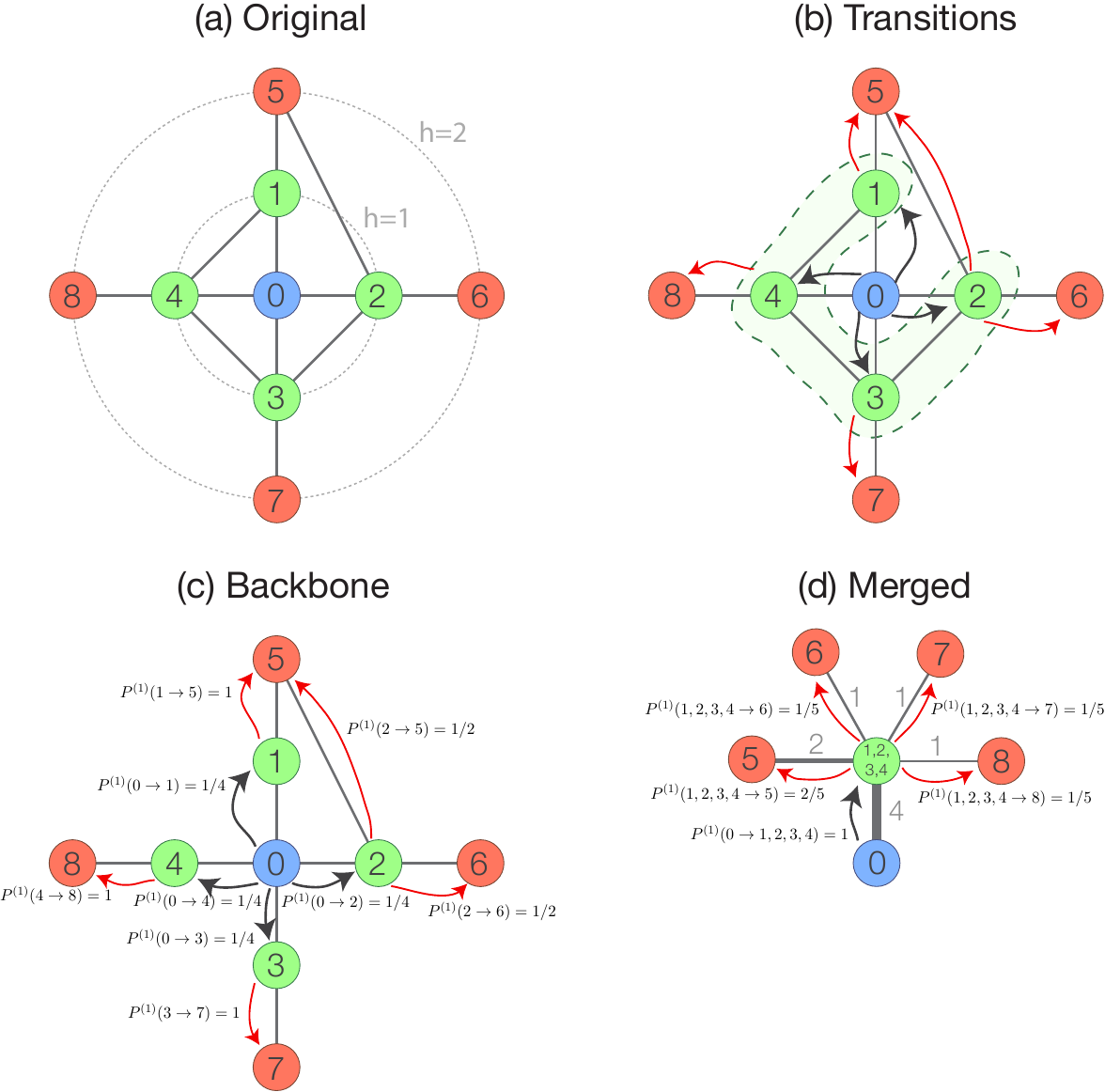}
  \caption{(Color online) Illustration of the concentric random walk dynamics. (a) Original 2-pattern, with concentric levels indicated by dashed lines. (b) One step transitions allowed for the random walk, indicated by arrows. Black and red arrows are for transitions departing from, respectively, the first and second concentric levels. Nodes inside the green dashed line form a connected group inside the first concentric level. Treating the connections between these nodes as infinite-cost or zero-cost is equivalent to representing the original pattern as, respectively, the (c) backbone and (d) merged patterns.}
  ~\label{f:exp_dyn}
  \end{center}
\end{figure*}

We define a \emph{l-pattern} centered at reference node $i$ as the subgraph of the original network containing all the nodes from $i$ to those in the $l$-th concentric level of $i$. In Fig.~\ref{f:exp_dyn}(a) we show an example of a 2-pattern and its concentric levels (indicated by dashed lines). In Fig.~\ref{f:exp_dyn}(b) the one-step transitions that the concentric random walk is allowed to move are indicated by arrows. In order to cover effectively the neighborhood of a node, after $h$ steps the agents of the concentric random walk should be able to arrive at all nodes of the $h$-th concentric level. Since allowing a transition between edges connecting nodes of the same concentric levels would make the coverage less effective, such connections are treated in a special manner. In fact, we define two possibilities for the transition over these edges, the infinite-cost and the zero-cost transitions. In the former case, the walker can never go through an edge connecting nodes at the same concentric level. In the zero-cost case, the walker can move freely inside the concentric level, that is, going from one node to another does not count as a step in the dynamics. These two cases can be exactly represented by two respective transformations of the network topology. For the infinite-cost transition, it suffices to remove all connections within the same concentric level, as shown in Fig.~\ref{f:exp_dyn}(c). We call the resulting network the \emph{backbone pattern} of the reference node. In the case of zero-cost transitions, connected nodes that are at the same concentric level are merged to form a single node, and the neighbors of the merged nodes are then connected to the resulting single node. Also, the number of connections made by the merged nodes with other concentric levels become the weight of the connection assigned to the new node. Fig.~\ref{f:exp_dyn}(d) shows the network obtained from the pattern in Fig.~\ref{f:exp_dyn}(a). We call the resulting network the \emph{merged pattern} of the reference node. For the backbone and merged patterns shown in Fig.~\ref{f:exp_dyn}, we indicate by arrows all possible one-step transitions for the concentric random walk, as well as their respective probabilities. In the remainder of this work, we will always use the terms backbone and merged to refer, respectively, to the infinite-cost and zero-cost transitions over edges within concentric levels. 
 
The backbone and merged representations can provide complementary information about the topological symmetry of the neighborhood of a node, therefore we define a symmetry measurement for each representation, which we call \emph{backbone symmetry} and \emph{merged symmetry}, respectively. Both symmetries are obtained by treating the backbone and merged patterns as follows. We represent as $P^{(h)}(i\rightarrow j)$ the probability that a walker undergoing a concentric random walk goes from node $i$ to node $j$ in $h$ steps. This allow us to define the following transition probabilities for nodes $j$ belonging to the set $\Gamma_i^{(h)}$,

\begin{figure*}[!tpb]
  \begin{center}
  \includegraphics[width=0.8\linewidth]{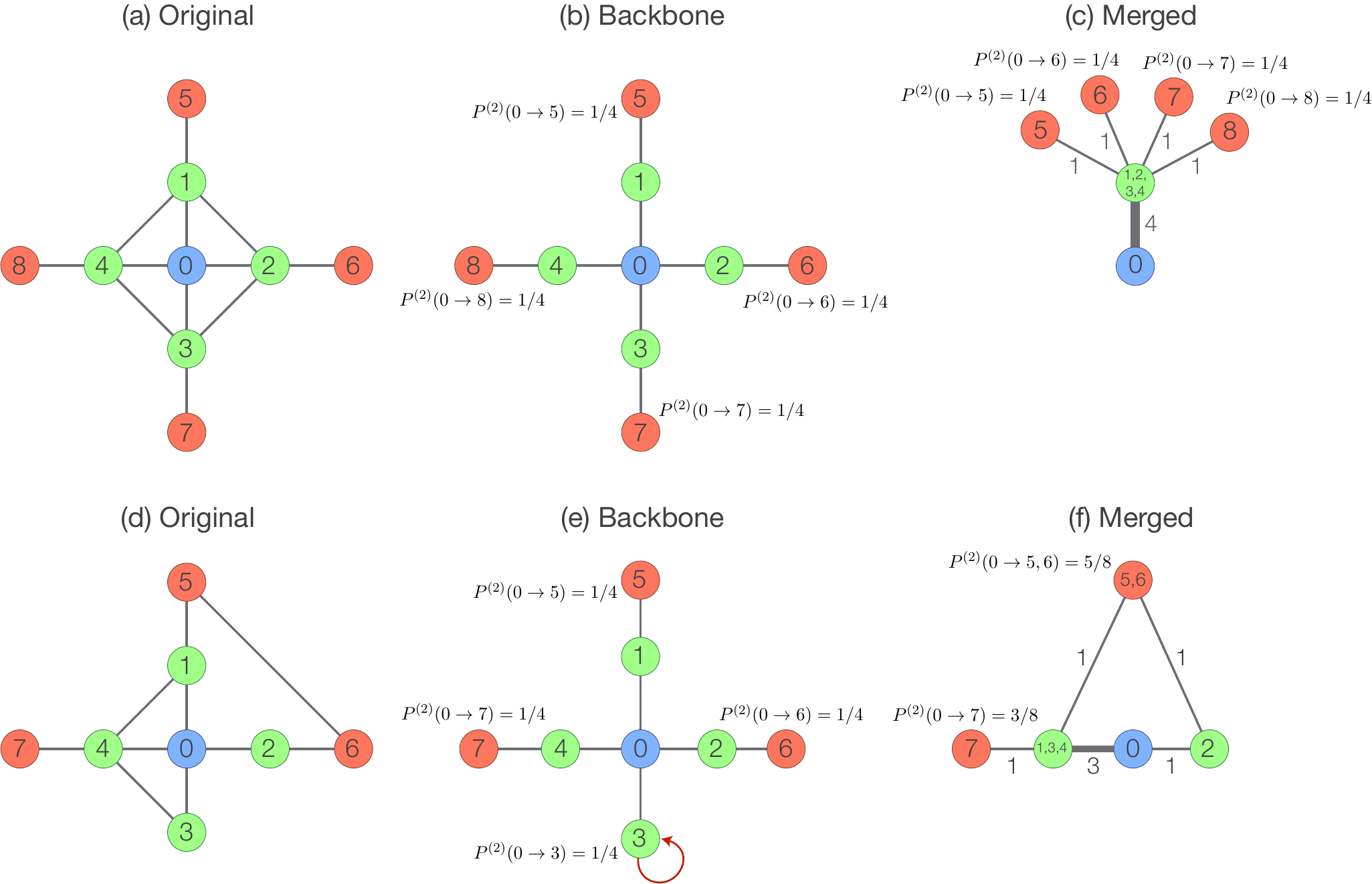} 
  \caption{(Color online) (a) Initial 2-pattern and its respective (b) backbone and (c) merged transformation. (d) Another example with a different 2-pattern and the respective (e) backbone and (f) merged patterns.}
  ~\label{f:exp_mea}
  \end{center}
\end{figure*}

\begin{equation}
P^{(h)}(i\rightarrow j|i\rightarrow k, \forall k \in \Gamma_i^{(h)})=\frac{P^{(h)}(i\rightarrow j)}{\sum\limits_{k\in\Gamma_i^{(h)}}P^{(h)}(i\rightarrow k)}.
\end{equation}
This expression can be understood as the probability of the walker being at node $j$, given that it arrived at some node of the $h$-th concentric level of node $i$. For simplicity's sake, we henceforth replace $P^{(h)}(i\rightarrow j|i\rightarrow k, \forall k \in \Gamma_i^{(h)})$ by $P^{(h)}(i\rightarrow j |i\rightarrow\Gamma_i^{(h)})$. We then calculate the Shannon entropy of the transition probabilities as

\begin{small}
\begin{align}
 \nonumber  H_i^{(h)}&=-\sum\limits_{j\in\Gamma_i^{(h)}}P^{(h)}(i\rightarrow j|i\rightarrow\Gamma_i^{(h)})\ln[P^{(h)}(i\rightarrow j|i\rightarrow\Gamma_i^{(h)})] \\ 
 &=-\sum\limits_{j\in\Gamma_i^{(h)}}\frac{P^{(h)}(i\rightarrow j)}{P^{(h)}(i\rightarrow \Gamma_i^{(h)})}\ln\left(\frac{P^{(h)}(i\rightarrow j)}{P^{(h)}(i\rightarrow \Gamma_i^{(h)})}\right).
\end{align}
\end{small}
Now defining $\xi_i^{(h)}$ as the set of reachable nodes from node $i$, for walks of length $h$ (i.e., $\xi_i^{(h)}=\{j|P^{(h)}(i\rightarrow j)>0\}$), we calculate the respective symmetry measurement as 

\begin{equation}
S_i^{(h)}=\frac{e^{H_i^{(h)}}}{|\xi_i^{(h)}|},
\label{eq:subgraph_symmetry}
\end{equation}
where $|X|$ is the cardinality of the set $X$. Therefore, the backbone and merged symmetry measurements, $\{Sb_i^{(h)}, Sm_i^{(h)}\}$, are assigned to the $h$-th concentric level of each node $i$ of the network. Note that we use the exponential of the entropy to have the number of reachable nodes as dimension \cite{viana2012effective}, which is more intuitive and compatible with $|\xi_i^{(h)}|$. As a consequence, both measurements of symmetry are conveniently bounded between 0 and 1. The normalization by $|\xi_i^{(h)}|$, instead of $|\Gamma_i^{(h)}|$, is needed because the concentric random walk can get trapped before arriving at the $h$-th concentric level. In such cases, the probabilities $P^{(h)}(i\rightarrow j |i\rightarrow\Gamma_i^{(h)})$ remain the same, but the pattern cannot be perfectly symmetric. Therefore, the normalization term $|\xi_i^{(h)}|$ guarantees that patterns containing trapping nodes will be less symmetric than similar patterns that do not contain such nodes. 

These local symmetry measurements are illustrated in the following example. For the 2-pattern shown in Fig.~\ref{f:exp_mea}(a), we calculate the two symmetry measurements for node 0 with respect to the second concentric level. First, we create the backbone and merged patterns, as shown in Figs.~\ref{f:exp_mea}(b) and (c). The backbone pattern is the original network without the connections within the first concentric level. For the merged pattern, nodes 1,2,3 and 4 are merged into a single node, and this new node connects to all neighbors of the merged nodes. Since four connections between concentric levels were merged into a single one, the new connection between node 0 and the merged node has weight 4. For the backbone pattern, $\Gamma_0^{(2)}=\{5,6,7,8\}$ and $\xi_0^{(2)}=\{5,6,7,8\}$. Therefore, the entropy of the transition probabilities is calculated as
\begin{align*}
 H_0^{(2)}&=-\sum\limits_{j\in\Gamma_0^{(2)}}\frac{P^{(2)}(0\rightarrow j)}{P^{(2)}(0\rightarrow \Gamma_0^{(2)})}\ln\left(\frac{P^{(2)}(0\rightarrow j)}{P^{(2)}(0\rightarrow \Gamma_0^{(2)})}\right) \\
 &=-4\frac{1/4}{1}\ln\left(\frac{1/4}{1}\right) \\
 &=\ln(4)
\end{align*}

The backbone symmetry is then given by
\begin{align*}
Sb_0^{(2)}&=\frac{e^{H_0^{(2)}}}{|\xi_0^{(2)}|} \\
         &=\frac{4}{4}=1.
\end{align*}

For the merged pattern, the calculation is analogous, but with the difference that the edge weights are incorporated into the transition probabilities. In the case of the specific example above, both measurements attain the maximum possible value, which is expected since the original pattern is symmetric. 

In Fig.~\ref{f:exp_mea}(d), we provide another example where the neighborhood of the node is not perfectly symmetric. The respective backbone and merged patterns are shown in Figs.~\ref{f:exp_mea}(e) and (f). Note that a self-loop is added to node 3 of the backbone pattern. This is done because when the walker reaches node 3 it becomes trapped in that node. The probability of being at node 3 is not considered when calculating $H_0^{(2)}$, since only the transition probabilities to nodes at the second concentric level are relevant. Nevertheless, the self-loop is still necessary because node 3 is in the set of nodes reachable from node 0. Therefore, the nodes in the second concentric level are $\Gamma_0^{(2)}=\{5,6,7\}$ and the reachable nodes are $\xi_0^{(2)}=\{3,5,6,7\}$. The remaining calculation is the same as above. For this case, the backbone symmetry is $Sb_0^{(2)}=0.75$ and the merged symmetry is $Sm_0^{(2)}=0.97$.

\section{Network Models}

\begin{figure*}[!tpb]
\begin{center}
\subfigure[][]{\includegraphics[width=0.49\linewidth]{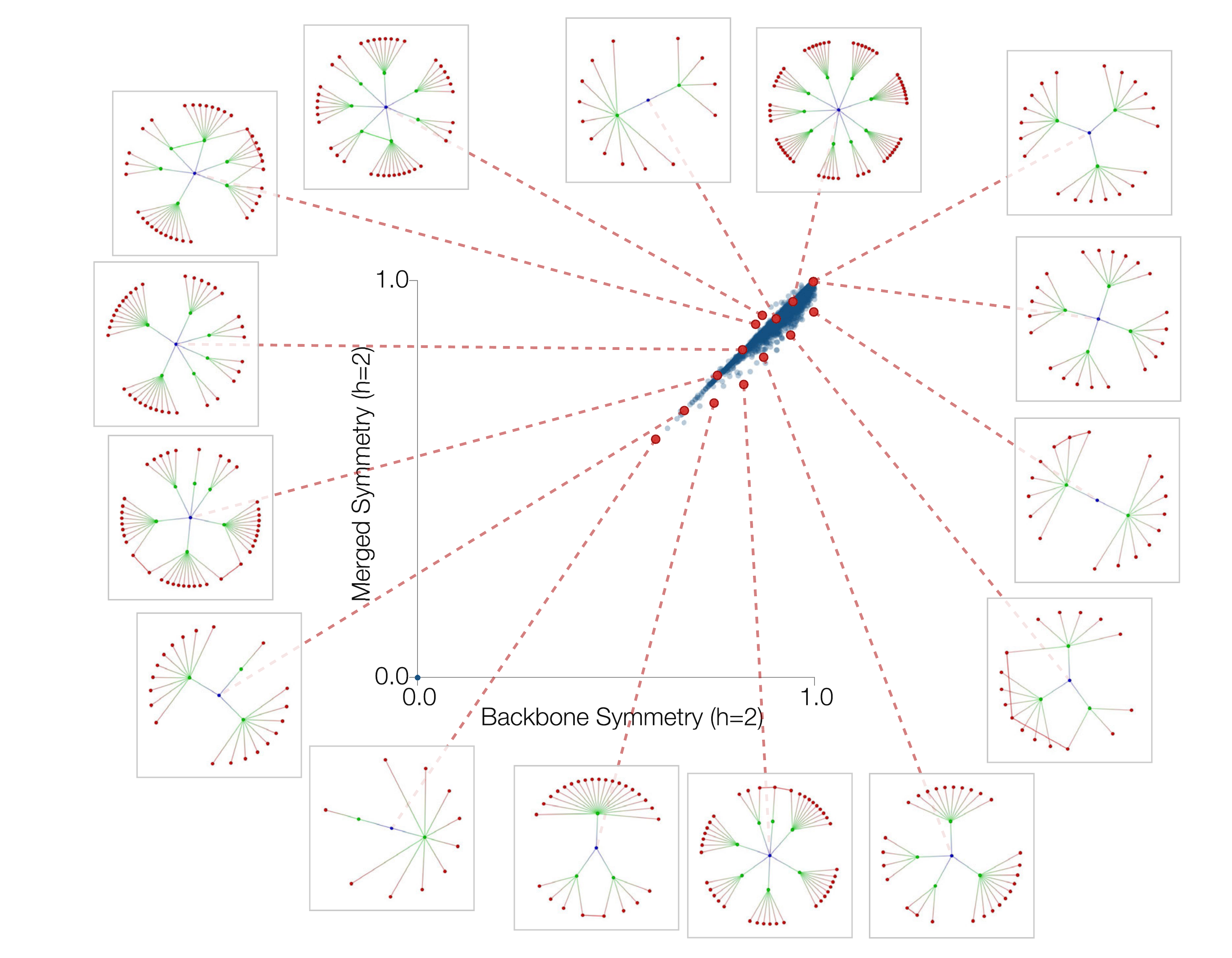} }
\subfigure[][]{\includegraphics[width=0.49\linewidth]{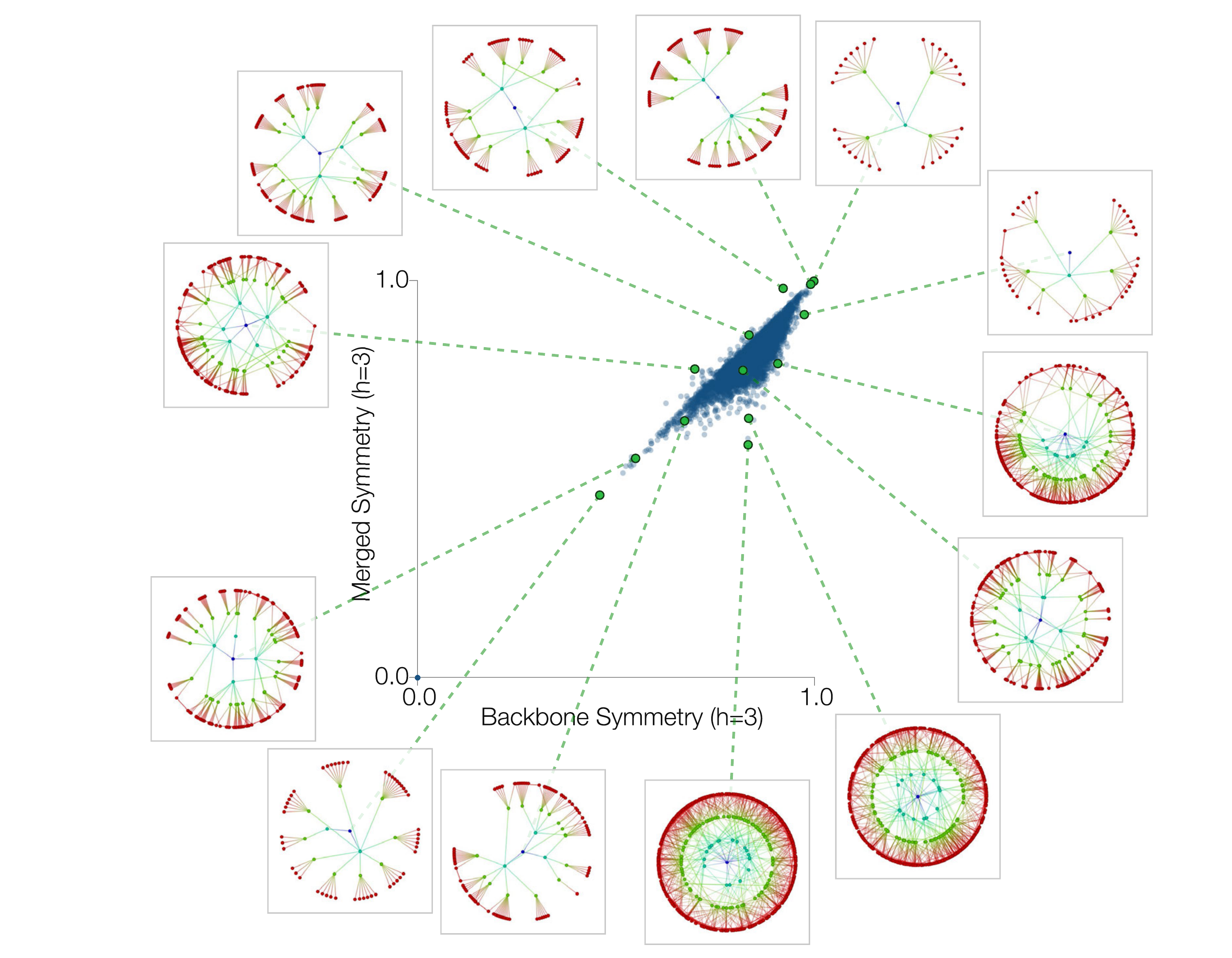} }
\end{center}
\caption{(Color online) Symmetries of the ER model for (a) $h=2$ and (b) $h=3$. The network has $N=5\times 10^3$ and average degree $\left\langle k \right\rangle = 6$.}
\label{f:ER_symm}
\end{figure*}

In this section we explore the potential of the two local symmetry measurements suggested in this article with respect to the characterization of several theoretical models of complex networks.  Let us start by characterizing the Erd\H{o}s-R\'enyi (ER) model in terms of the backbone and merged symmetries $\{Sb_i^{(h)}, Sm_i^{(h)}\}$. In Fig.~\ref{f:ER_symm} we show the scatterplots of $Sb_i^{(h)}$ as a function of $Sm_i^{(h)}$ for $h=2$ and $3$, together with some patterns with their corresponding symmetry values. For $h=2$ (Fig.~\ref{f:ER_symm} (a) ), we see that there is high density of subgraphs occurring for values of $Sb_i^{(2)}$ close to $Sm_i^{(2)}$, with a great homogeneity of observed patterns. Note, also, that most of these patterns are tree-like, i.e., without the presence of triangles. This pattern is frequently found in uncorrelated networks \cite{Newman010:book}. Therefore, for low values of $h$, the ER model yields highly symmetric patterns, essentially as a consequence of the aforementioned tree-like characteristic of the network, which is also related to its relative simplicity/regularity (in the sense that the overall connectivity can be well described in terms of the average degree). 

\begin{figure*}[!tpb]
\begin{center}
\subfigure[][]{\includegraphics[width=0.49\linewidth]{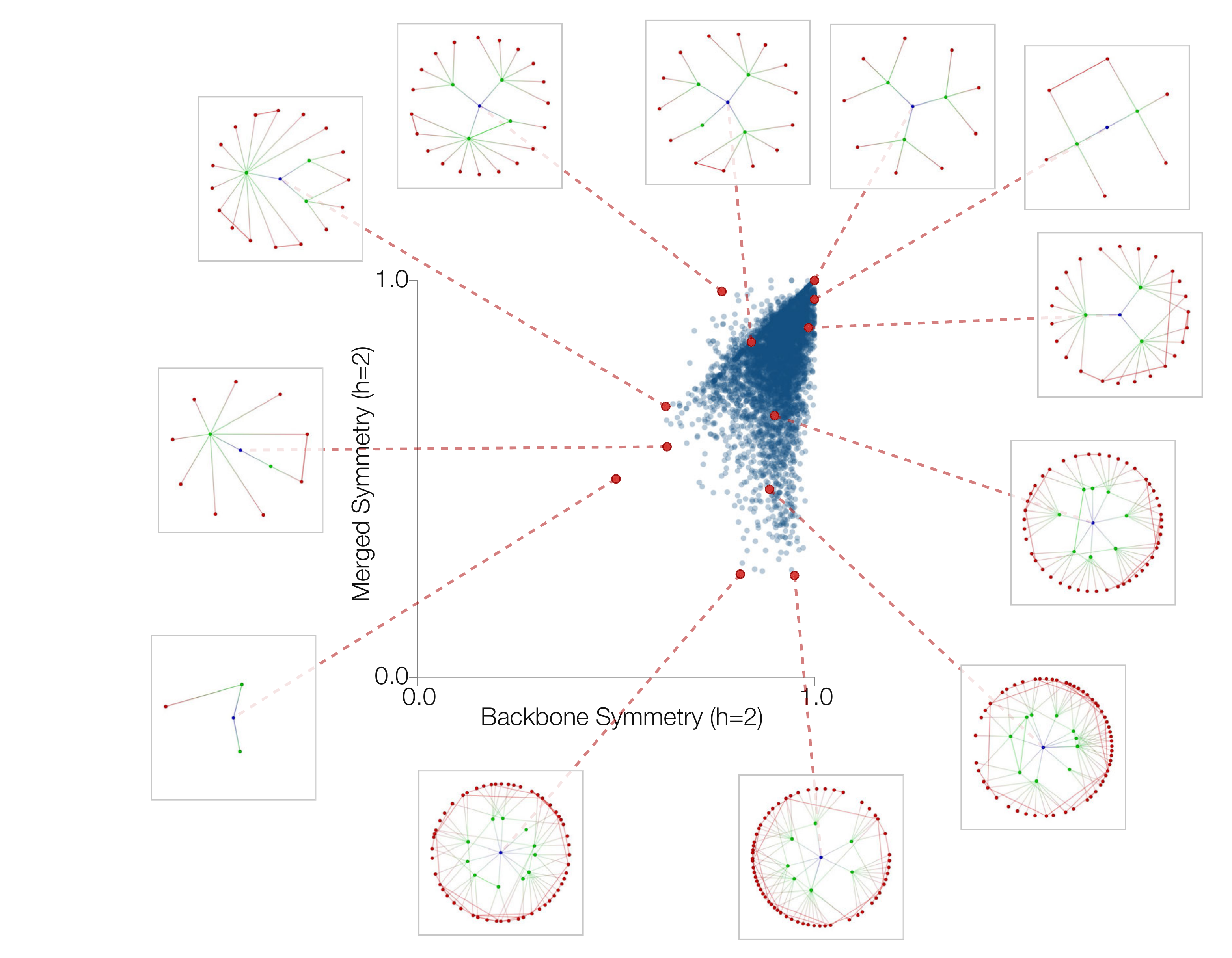} }
\subfigure[][]{\includegraphics[width=0.49\linewidth]{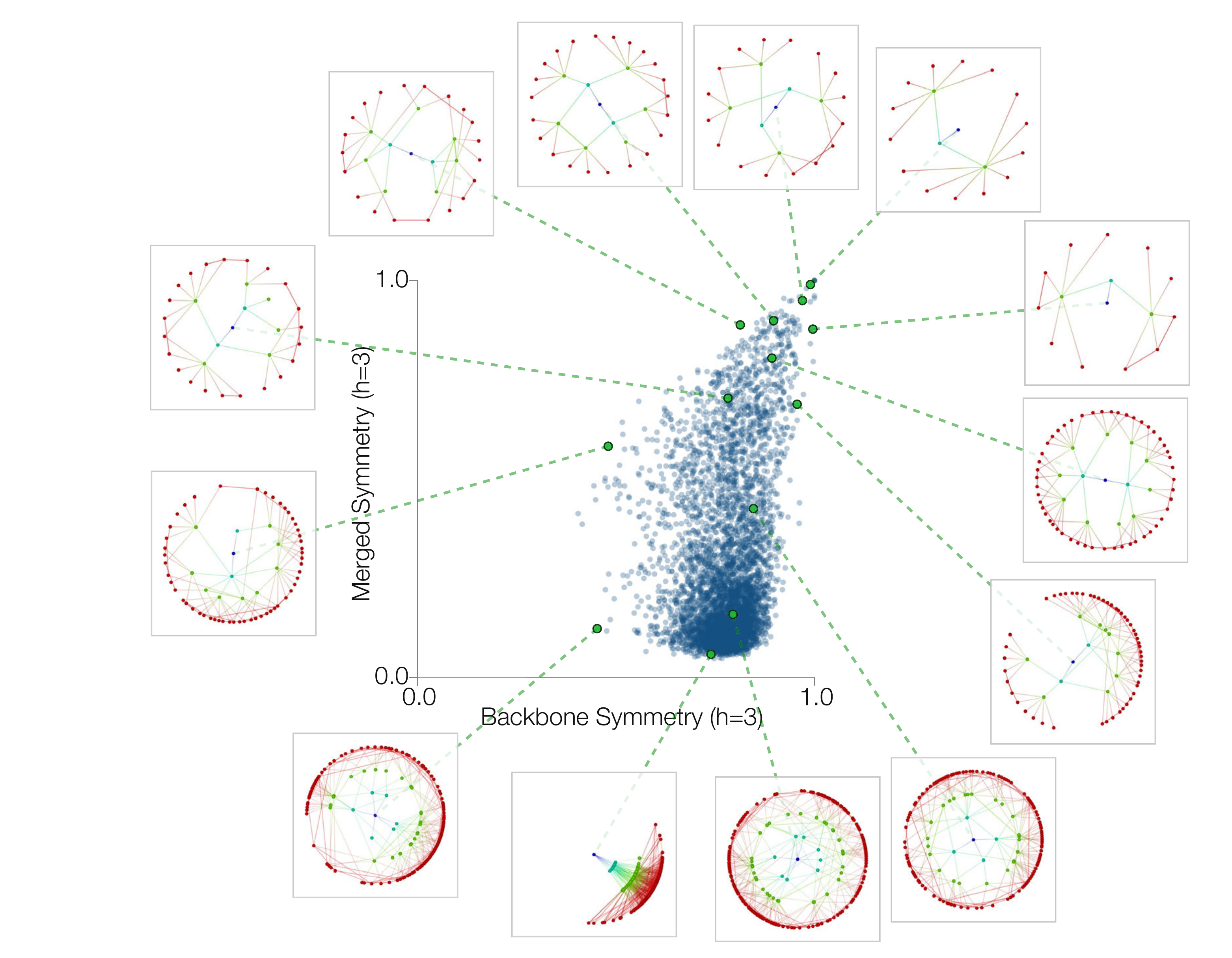} }
\end{center}
\caption{(Color online) Symmetries of the GEO model for (a) $h=2$ and (b) $h=3$. The network has $N=5\times 10^3$ and average degree $\left\langle k \right\rangle = 6$.}
\label{f:GEO_symm}
\end{figure*}

The above scenario changes as we increase the value of $h$. For $h=3$ (Fig.~\ref{f:ER_symm} (b)), the symmetry values are smaller than in the former case. This effect is stronger for the merged symmetry. This can be explained by the small-world characteristic of the ER model. Since these networks have average shortest path lengths of order $\ell \sim \ln N$, with a few steps all nodes can be reached (given that the networks contain only one connected component), becoming highly unlikely that a symmetric structure is obtained. Any eventual connection between the same level will not change the backbone symmetry, but will strongly impact on the merged symmetry. Since for larger $h$ the probability of having a connection on the same level is larger, the merged symmetry values tend to decrease. Note in Fig.~\ref{f:ER_symm} that patterns below the line $S_m^{(2)}=S_b^{(2)}$ often present a higher number of connections (see Fig.~S1). 

Next we turn our analysis to spatial random graphs to study spatial effects on the symmetry of patterns. First we consider the random geometric graphs (GEO) whose nodes are connected if they are separated by distances smaller than a certain distance~\cite{dall2002random}. As we can see in Fig.~\ref{f:GEO_symm}, the GEO network behaves similarly to the ER model, though with more accentuated decrease in the merged symmetry as we increase the value of $h$. Unlike the ER model, GEO networks do not have local tree-like structures, but rather a high number of triangles implied by connecting all nodes within a certain radius. The increase in the clustering coefficient leads, therefore, to more connections between nodes at the same concentric level. This is reflected by the larger dispersion in the values of the merged symmetry for $h=2$ in Fig.~\ref{f:GEO_symm}(a). Moreover, for $h=3$ (Fig.~\ref{f:GEO_symm}(b)), we observe a striking difference between the GEO networks and the ER structures. Although perfectly symmetric patterns can still be observed, the symmetry of the merged patterns decrease significantly, suffering even more the effects of high clustering. 

\begin{figure*}[!tpb]
  \begin{center}
  \includegraphics[width=0.6\linewidth]{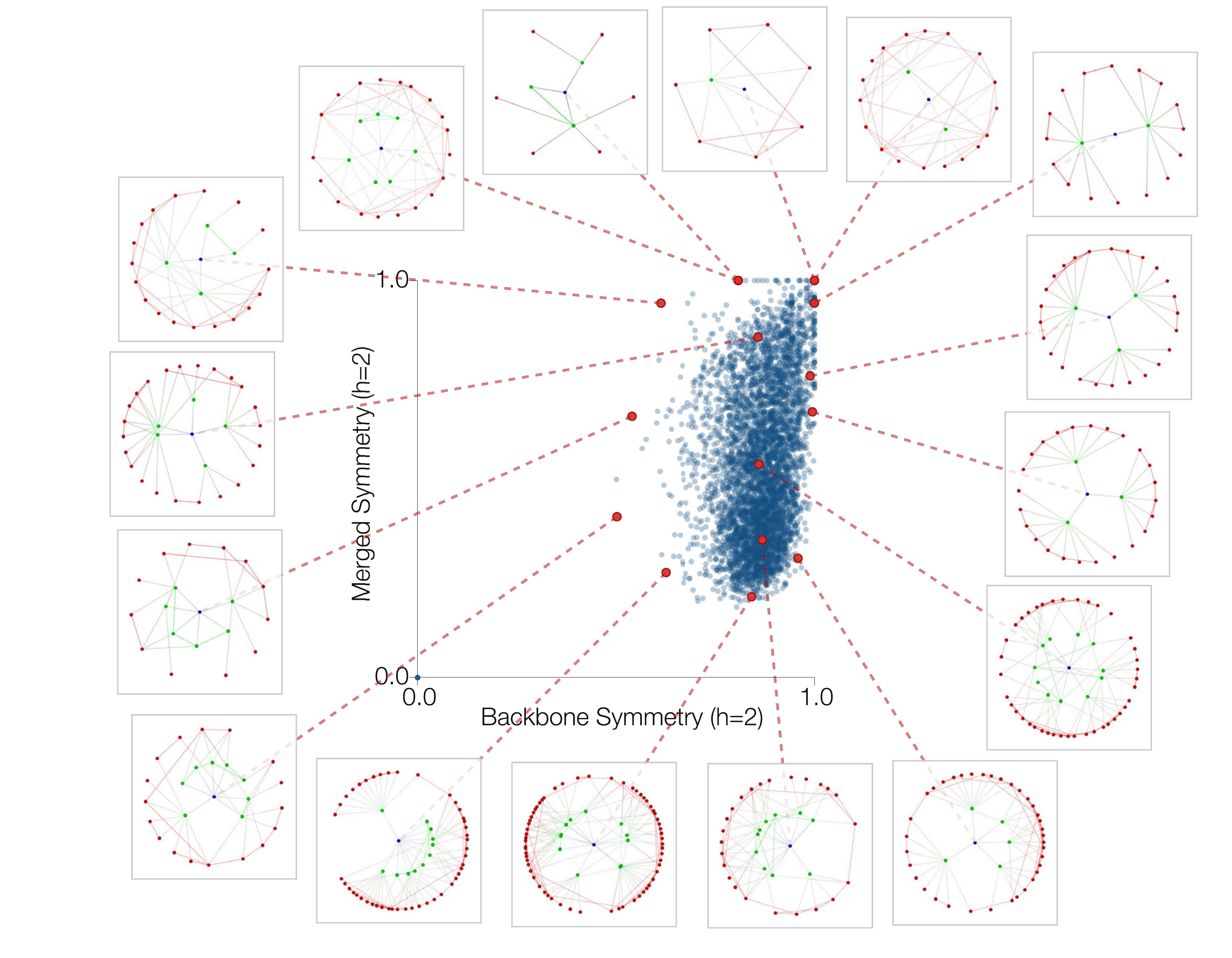} 
  \caption{(Color online) Symmetries for a WAX network with $N=5\times 10^3$ and $\left\langle k \right\rangle=6$.}
  ~\label{f:Wax}
  \end{center}
\end{figure*}

Another example of a spatial random graph, namely the Waxman model (WAX) ~\cite{waxman1988routing,barthelemy2011spatial}, is shown in Fig.~\ref{f:Wax}. Interestingly, both symmetry measurements tend to be smaller for this structure compared to the values obtained for GEO. It is worth noting that there is a non-vanishing probability of observing long-distance connections in the WAX model, that might explain the loss of backbone symmetry for larger values of $h$. 

\begin{figure*}[!tpb]
\begin{center}
\subfigure[][]{\includegraphics[width=0.49\linewidth]{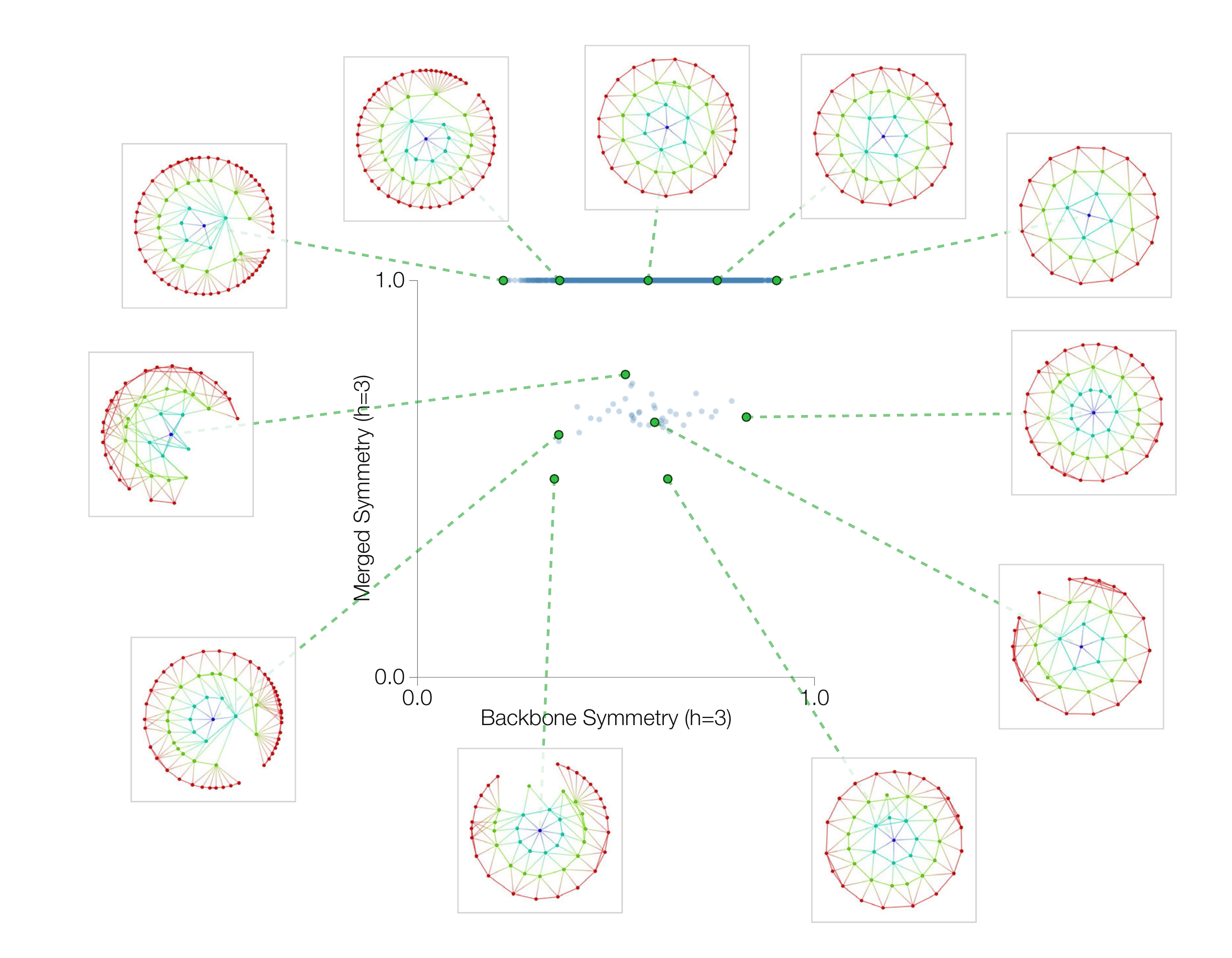} }
\subfigure[][]{\includegraphics[width=0.49\linewidth]{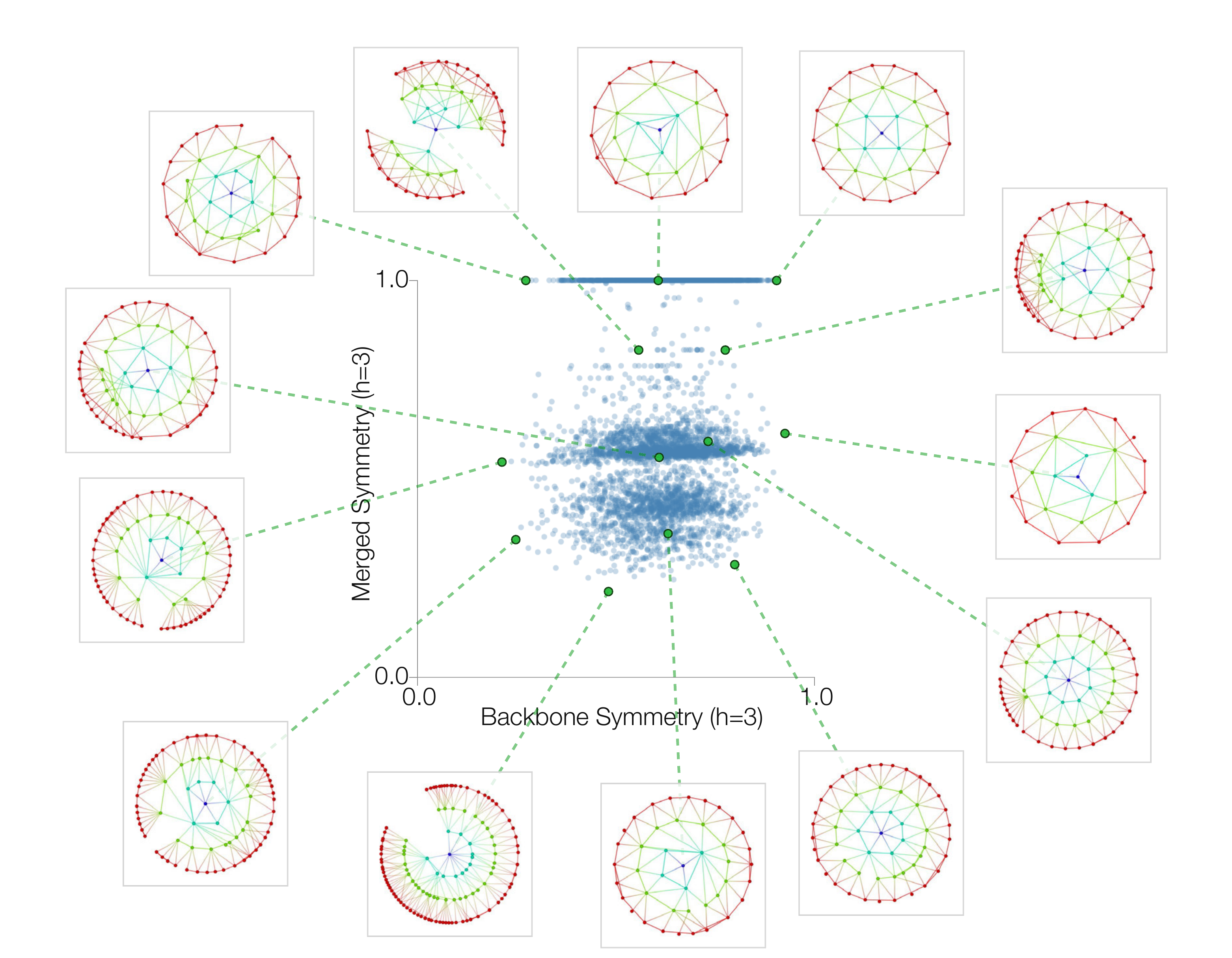} }
\end{center}
\caption{(Color online) Symmetries for $h=3$ of the (a) VOR model and (b) RVOR model with probability $0.001$ of random rewiring. Each network has $N=5\times 10^3$ and average degree $\left\langle k \right\rangle = 6$.}
\label{f:Voronoi_symm}
\end{figure*}

Voronoi networks~\cite{barthelemy2011spatial, estradaBook2012, costa2004voronoi} (VOR) are typically obtained from a given spatial distribution of nodes in a metric space, with pairs of nodes being interconnected whenever their Voronoi cells touch one another. Therefore, no edge crossings are allowed and the network is necessarily planar. This model is particularly interesting to examine symmetries when strong spatial restrictions are imposed to the system (e.g. road networks). Fig.~\ref{f:Voronoi_symm}(a) illustrates the backbone and merged symmetries for $h=3$, as well as some representative patterns. The strong spatial restriction generates highly symmetric patterns even though presenting large fluctuations on the connectivity between concentric levels for the backbone patterns. The effects of adding a few shortcuts to the network is also interesting to investigate, since this removes the spatial adjacency imposed in the network construction. Fig.~\ref{f:Voronoi_symm}(b) presents the results obtained for a rewired Voronoi network (RVOR), generated by randomly selecting links in the previous network and placing them between a randomly chosen pair of nodes (the rewiring probability is $0.001$). It is clear that the few added shortcuts are enough to considerably reduce the overall symmetry of the network patterns, thus illustrating the special role played by spatial restrictions on the symmetry.

\begin{figure*}[!tpb]
  \begin{center}
  \includegraphics[width=0.6\linewidth]{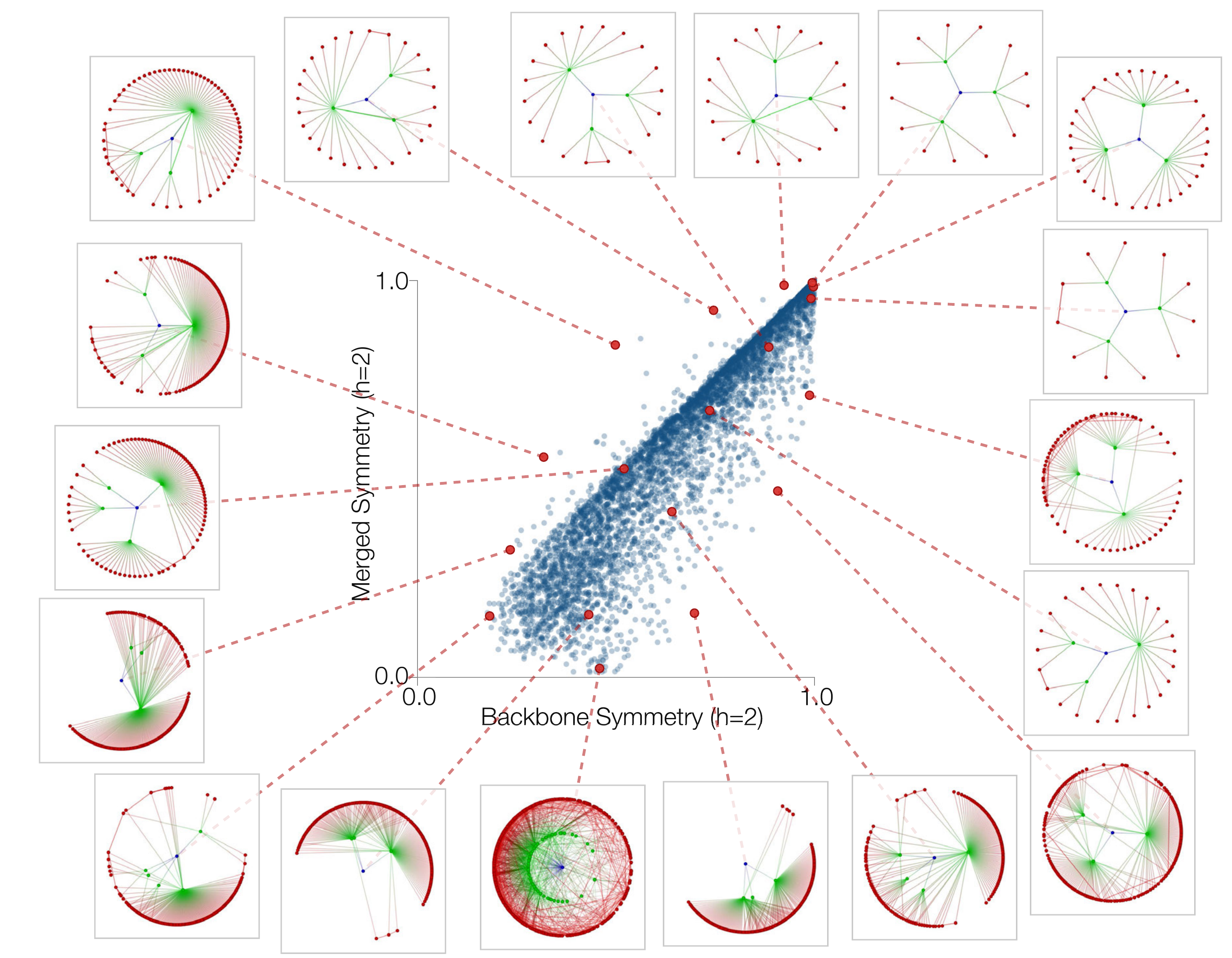}
  \caption{(Color online) Symmetries for $h=2$ of a BA network with $N=5\times 10^3$ and $\left\langle k \right\rangle=6$.}
  ~\label{f:BA}
  \end{center}
\end{figure*}

The Barab\'asi-Albert (BA) model is also considered. We note in Fig.~\ref{f:BA} that the most symmetric patterns are those centered at nodes with smaller degree. On the other hand, hubs have lower values of merged symmetry, as high degrees tend to imply larger probability of reaching non-symmetric regions in the network.

It is also interesting to verify how the symmetry values correlate with the node degree. Scatterplots are included in the Supplementary Material for the second, third and fourth concentric levels. The symbols are colored according to the degree of the reference node. For all models (shown in Figs.~S1, S2, S3, S4, S5 and S6), except BA, nodes with higher degrees tend to present lower symmetry values, specially for the merged symmetry. This happens because nodes with larger degrees commonly have more nodes at their initial concentric levels. Since it is less probable that patterns with a large number of nodes remains symmetric, such nodes tend to have lower symmetry. 

\begin{figure*}[!tpb]
\begin{center}
\subfigure[][]{\includegraphics[width=0.44\linewidth]{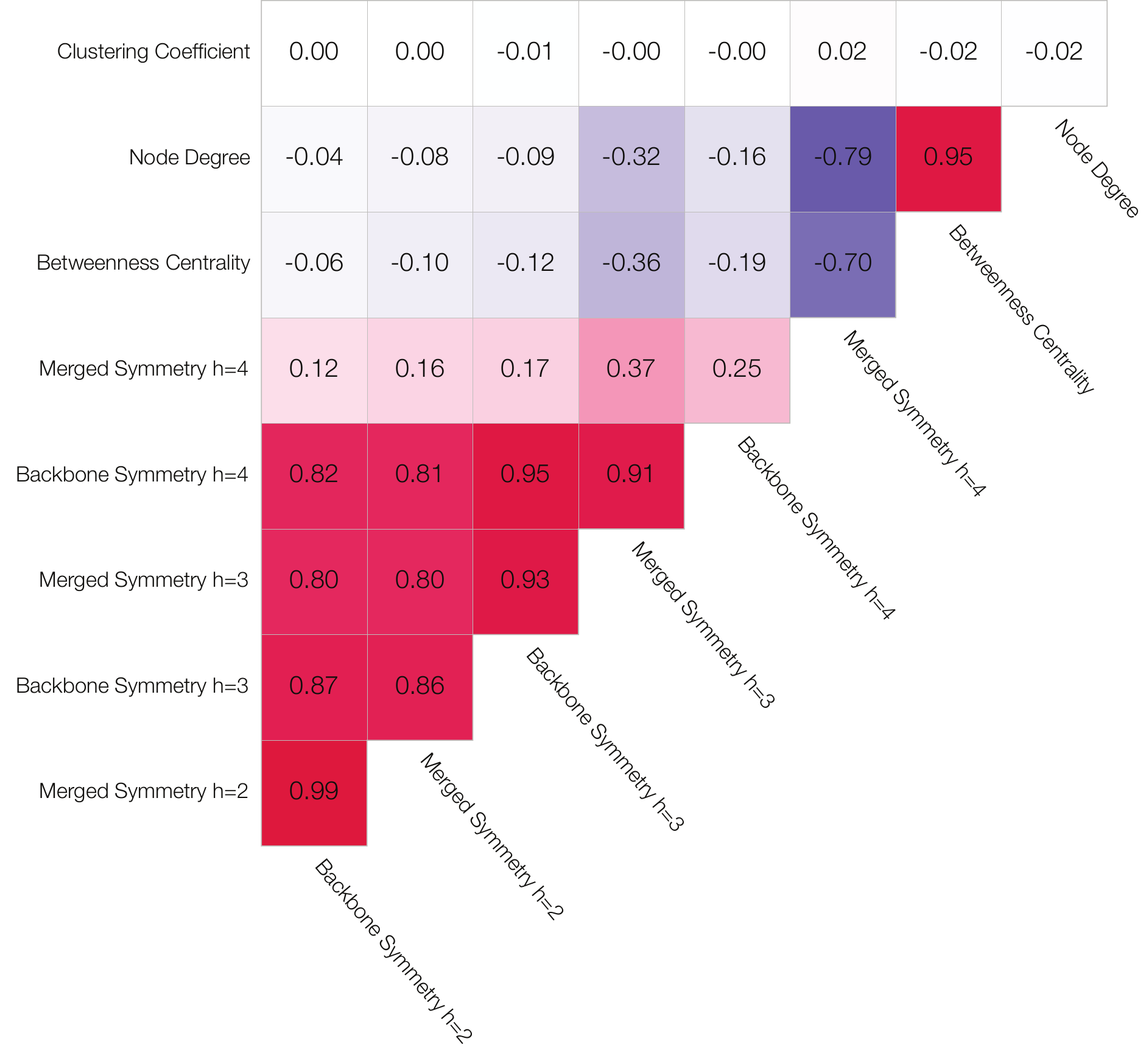} }
\subfigure[][]{\includegraphics[width=0.44\linewidth]{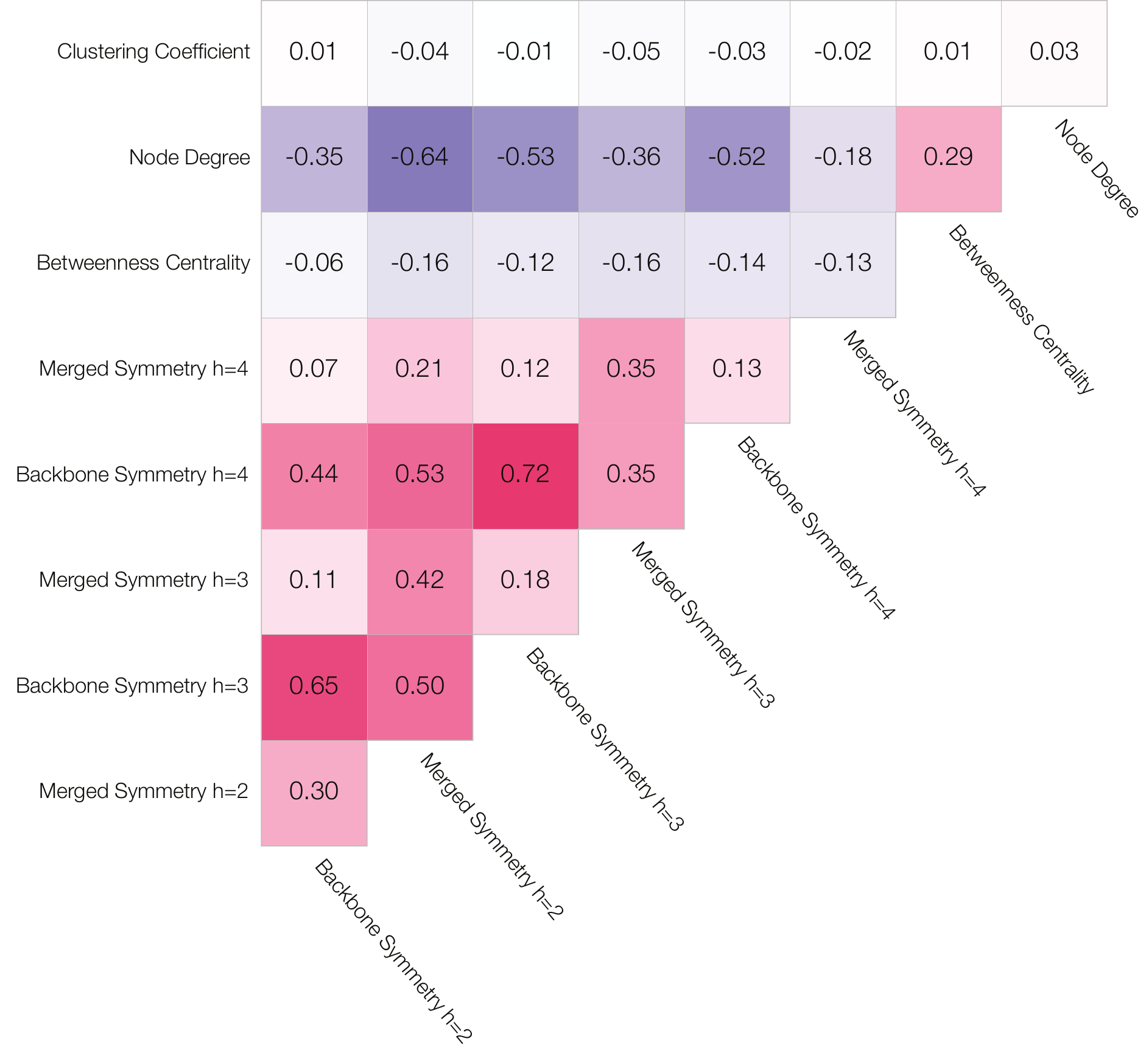} }
\end{center}
\caption{(Color online) Correlations between the symmetry measurements proposed here with some topological properties for (a) ER and (b) WAX model.}
\label{f:correlations_model}
\end{figure*}

We also investigate the correlations between the symmetry measurements and between these and three traditional topology measurements, namely the degree, clustering and betweenness centrality. The Pearson correlation values, organized as a matrix, obtained for the ER and WAX networks are shown in Figs.~\ref{f:correlations_model}(a) and (b), respectively. The results are given only for two models because the correlation matrix of the BA case are similar to those of the ER model, and the matrix of the GEO network is similar to the WAX case. The backbone symmetry for $h=2,3,4$ and merged symmetry for $h=2,3$ show strong correlations among themselves for the ER model. This is a consequence of the high regularity of this model, where the same neighborhood patterns appear at different concentric levels. The WAX model shows less correlation between the symmetry measurements, which might be explained by the large variability of patterns which is characteristic of this model. Also, there is low correlation between the symmetry measurements and the traditional ones, which indicates that the new measurements defined in the current work provide complementary information about the network structure that is not contained in the other measurements.

\section{Real-world Networks}

The potential of the proposed symmetries for characterizing the local structure of some real-world networks is investigated in the following.

\begin{figure*}[!tpb]
  \begin{center}
  \includegraphics[width=0.6\linewidth]{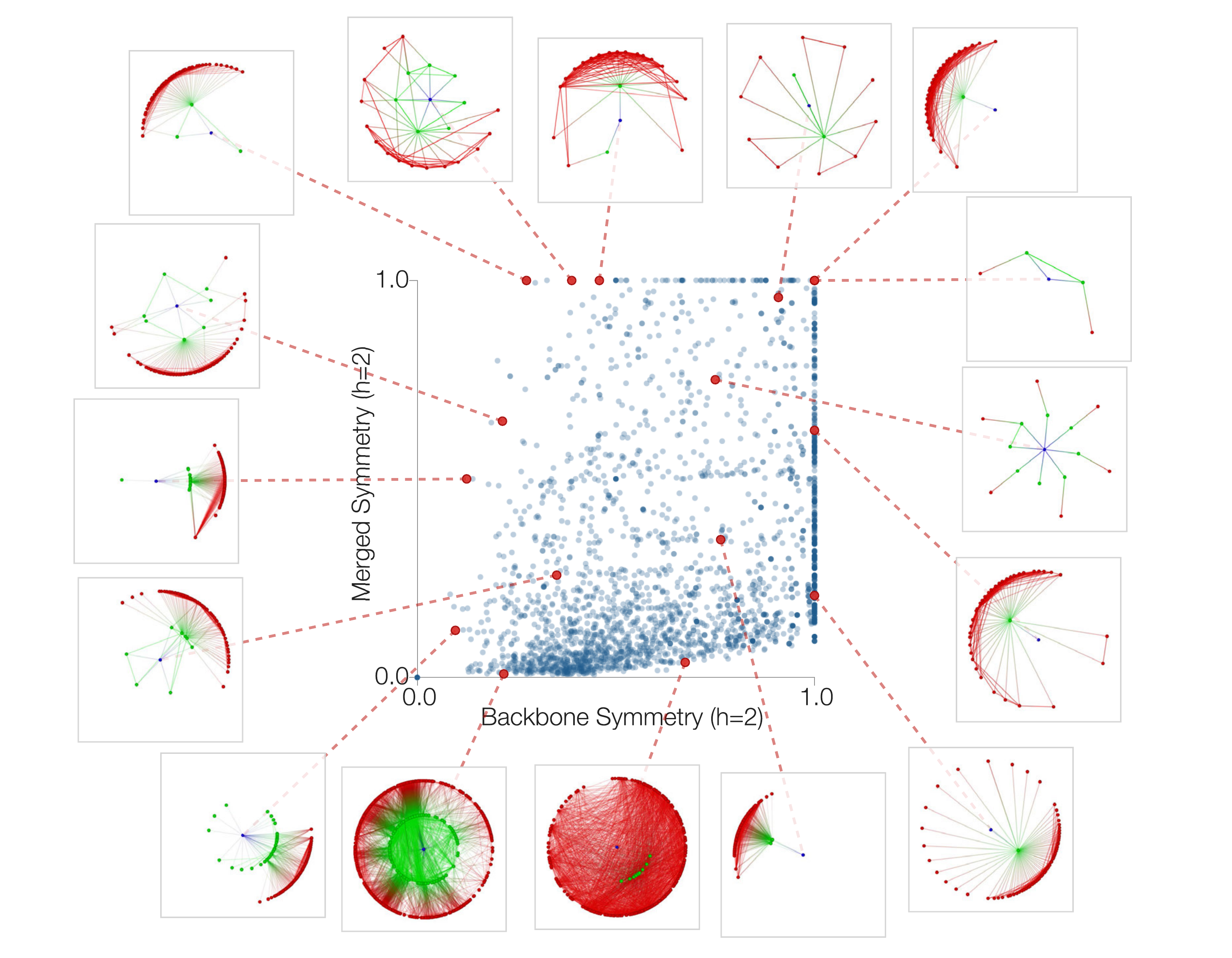} 
  \caption{(Color online) Symmetries of the patterns for the airport network for $h=2$.}
  ~\label{f:airport}
  \end{center}
\end{figure*}

Fig.~\ref{f:airport} shows the concentric patterns of the world-wide airports network ($N=2940$ and $\langle k\rangle \approx 21$) in which pairs of airports are connected if they share a route \footnote{The airport and route data were obtained from http://sourceforge.net/p/openflights/code/HEAD/tree/openflights/data/}. Interestingly, a whole different scenario is observed than that obtained for the network models. For instance, we see that the values of backbone and merged symmetry are much more distributed in the airport network than in any of the considered network model. Another interesting trend is that a significant number of patterns have maximum values for one of the symmetries. Those with $S_b^{(2)}=1$ correspond to patterns centered at nodes with low degrees connected to hubs, yielding tree-like patterns. These nodes represent small airports connected to larger airports that naturally have many connections. The patterns lying at $S_m^{(2)}=1$ correspond to highly clustered patterns, mainly composed by links at the same concentric level. 

\begin{figure*}[!tpb]
  \begin{center}
  \includegraphics[width=0.6\linewidth]{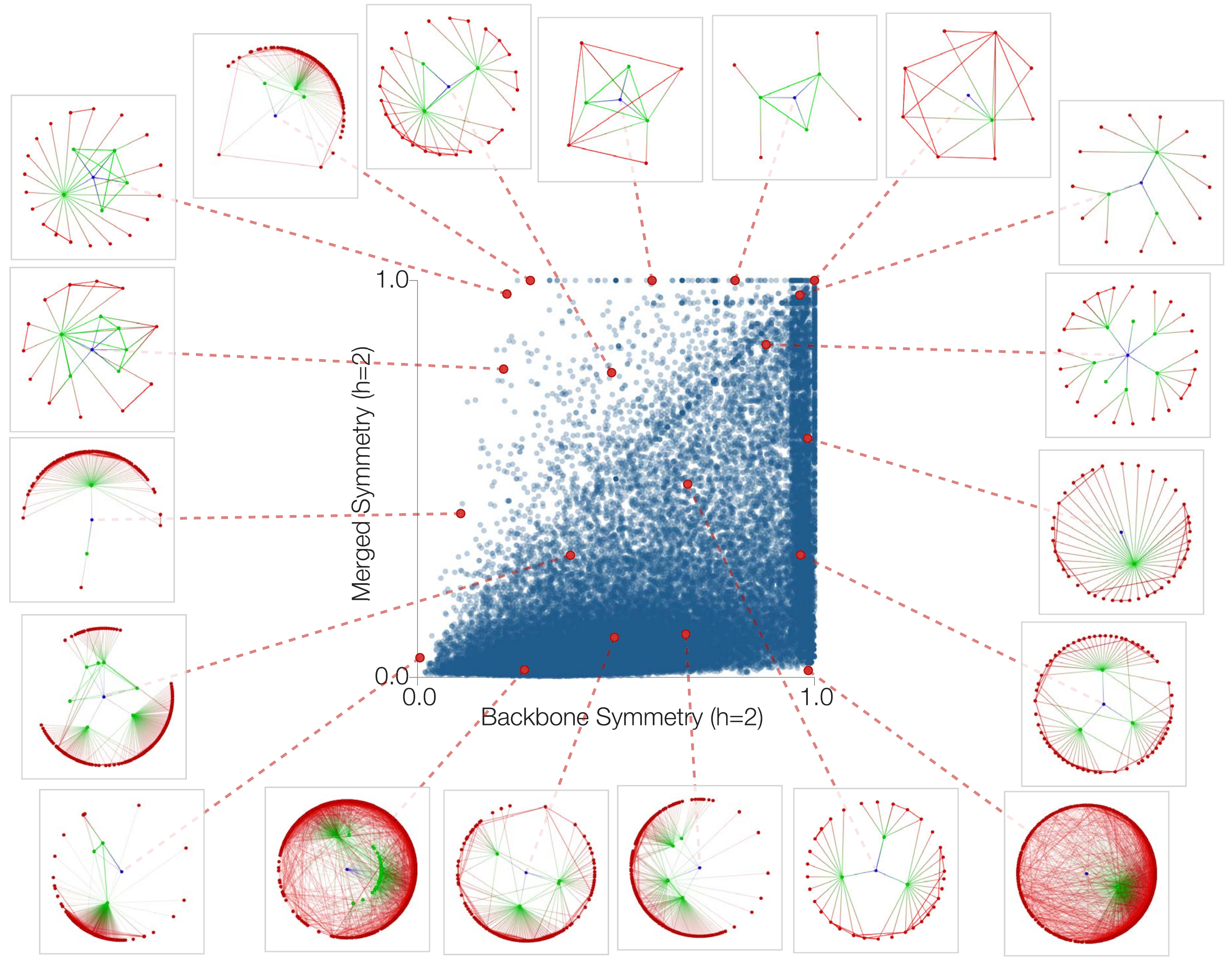} 
  \caption{(Color online) Symmetries of the patterns for the scientific areas of the Wikipedia network for $h=2$.}
  ~\label{f:Wiki}
  \end{center}
\end{figure*}

A similar behavior is observed for the scientific areas of the Wikipedia network ($N=45876$ and $\langle k\rangle \approx 5.8$). Here we consider that two pages in Wikipedia are connected if one is cited by the other, ignoring the direction of the citation. Again, as we can see in Fig.~\ref{f:Wiki}, the dispersion in the symmetry space contrasts to the patterns obtained or the model cases. 

\begin{figure*}[!tpb]
  \begin{center}
  \subfigure[][]{\includegraphics[width=0.49\linewidth]{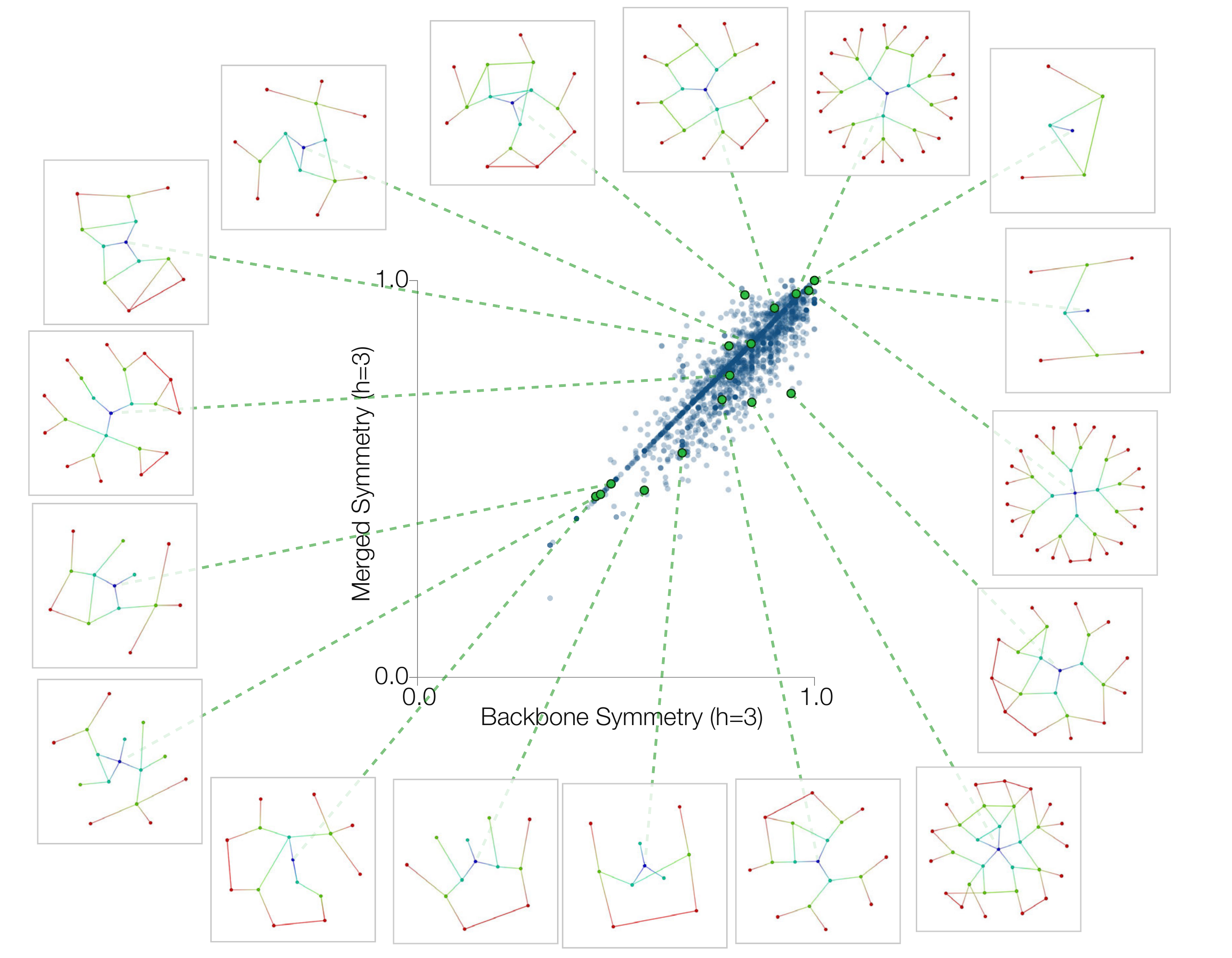}}
  \subfigure[][]{\includegraphics[width=0.49\linewidth]{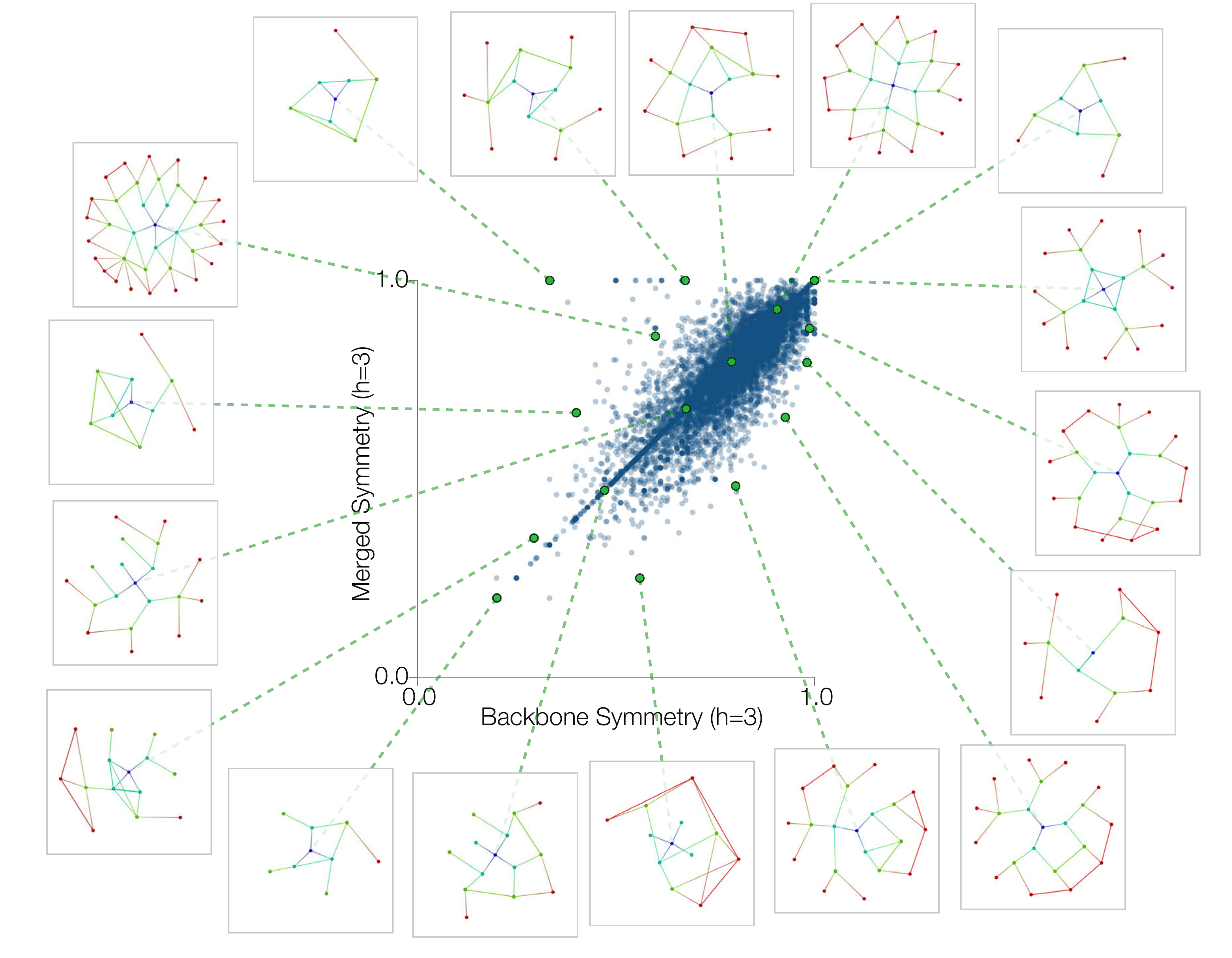}}
  \caption{(Color online) Symmetries of the patterns for the Oldenburg (a) and San Joaquin (b) street networks for $h=3$.}
  ~\label{f:Oldenburg_n_San_Joaquin}
  \end{center}
\end{figure*}

Remarkably, the contrast between real-world networks and random graph models is not observed for the real-world spatial networks considered here, namely, the street networks of Oldenburg ($N=14503$ and $\langle k\rangle \approx 2.8$) and San Joaquin~\cite{Brinkhoff2002} ($N=2873$ and $\langle k\rangle \approx 2.6$), see Fig.~\ref{f:Oldenburg_n_San_Joaquin}. In fact, the symmetry plot is similar to those observed for the ER model. Interestingly, even though the two cities have different street plannings \cite{Lacasa2013}, both present similar dependence between $Sb_i^{(h)}$ and $Sm_i^{(h)}$.

The Supplementary Material contains the scatterplots of the symmetry measurements colored according to the node degree. The airport (shown in Fig.~S7), Wikipedia (shown in Fig.~S8) and street networks (shown in Figs.~S9 and S10) show the expected behavior of smaller symmetry values when the degree is increased. The only noticeable exceptions to this trend are the airport and Wikipedia networks for $h=4$.

\begin{figure*}[!tpb]
\begin{center}
\subfigure[][]{\includegraphics[width=0.44\linewidth]{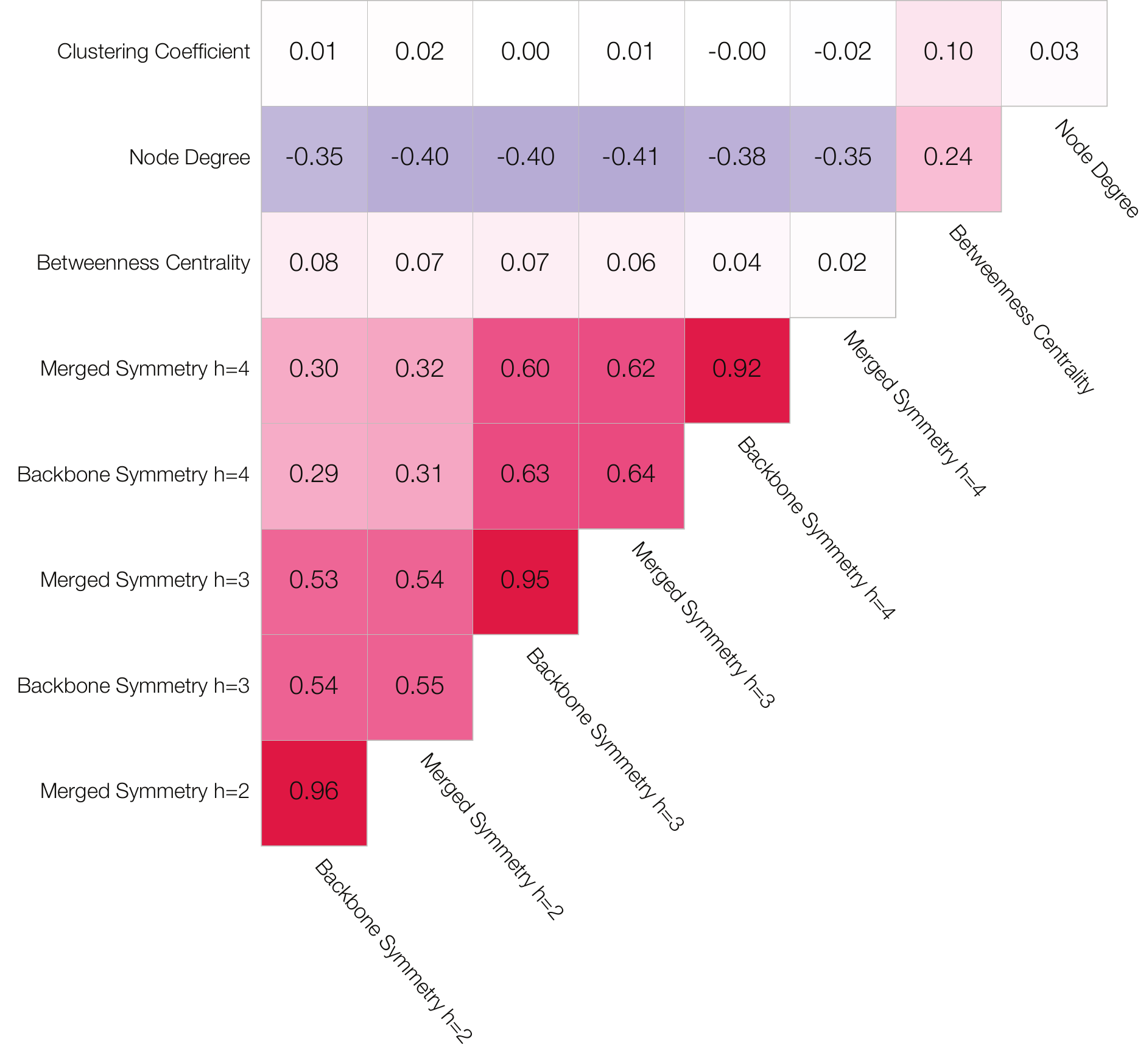} }
\subfigure[][]{\includegraphics[width=0.44\linewidth]{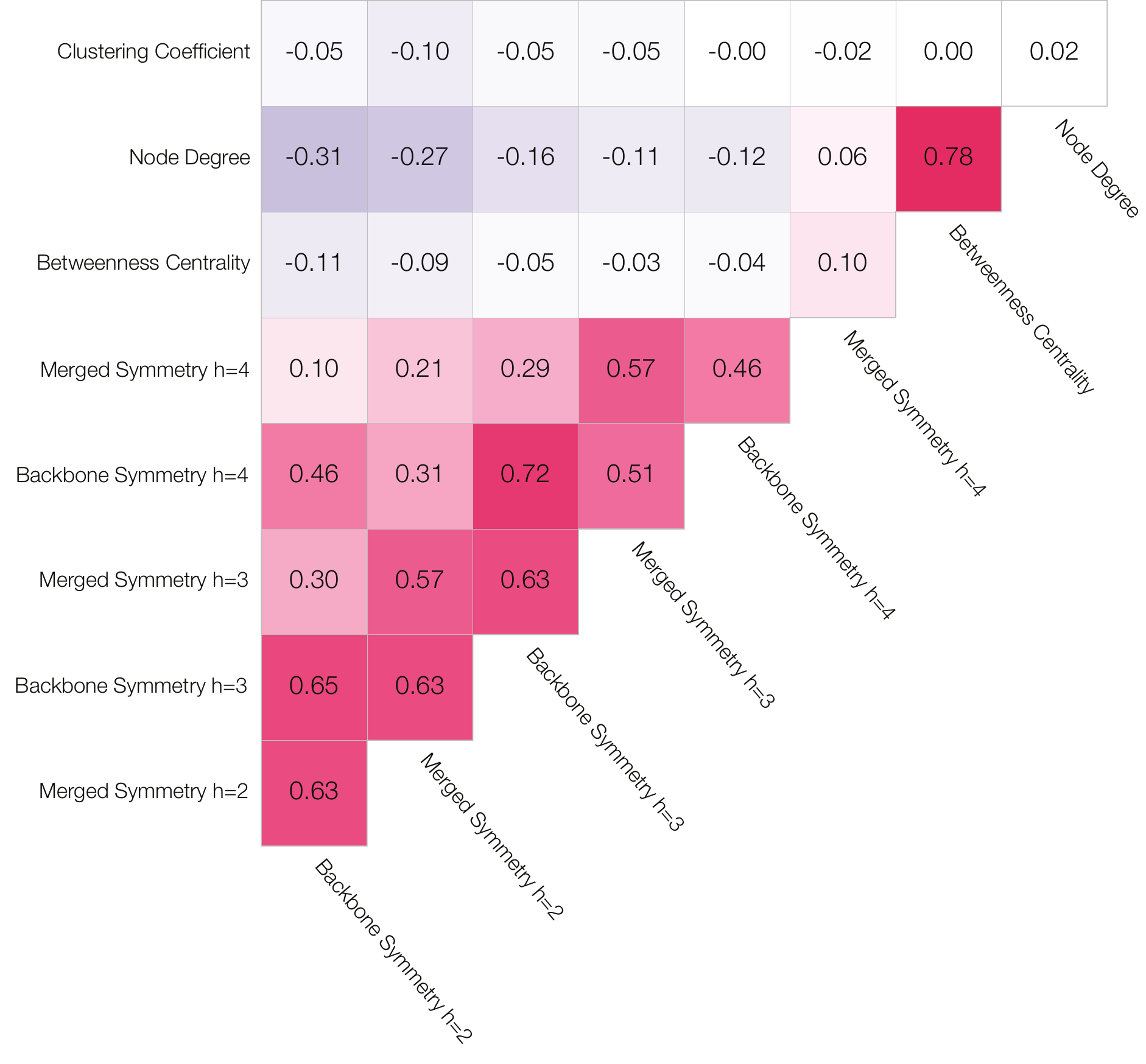} }
\end{center}
\caption{(Color online) Correlations between the symmetry measurements proposed here with some traditional topological properties for (a) the Oldenburg street network and (b) the Wikipedia citation network.}
\label{f:correlations_real}
\end{figure*}

The matrices with the Pearson correlation coefficients between the symmetry indices and three traditional topological measurements are shown in Figs.~\ref{f:correlations_real}(a) and (b) for the Oldenburg street network and the Wikipedia network, respectively. We note that the correlation matrix for San Joaquin is similar to the matrix calculated for Oldenburg, and the correlations for the Wikipedia are similar to those in the airport network. The Oldenburg network, as expected, has large correlations between backbone and merged symmetries calculated at the same respective level. For both networks, the correlations between the symmetry measurements of distinct concentric levels are low, indicating that these measurements, taken along the concentric levels, provide complementary information about the networks organization. Also, the correlations between the traditional measurements and the symmetry are particularly low. 

\section{Statistical analysis}
\begin{figure*}[!tpb]
\begin{center}
\subfigure[][Symmetry measurements]{\includegraphics[height=8cm]{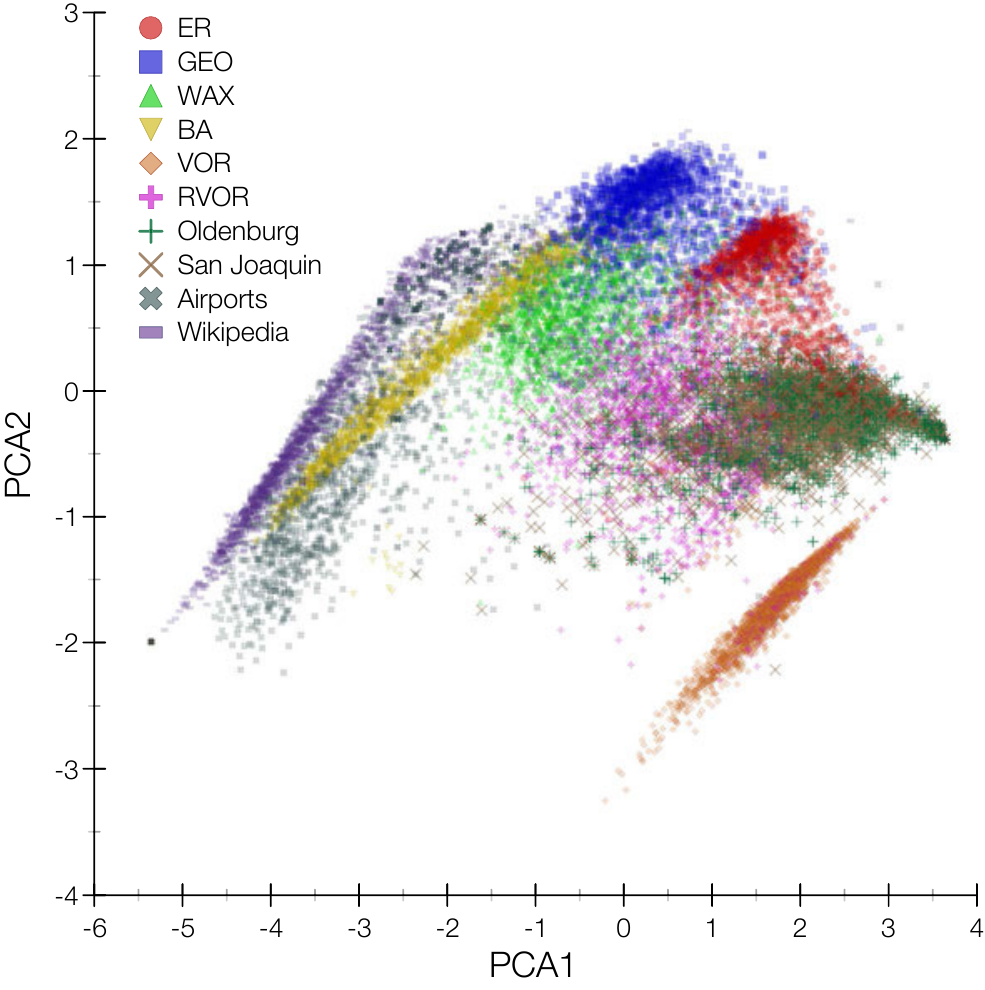} }
\subfigure[][Concentric measurements]{\includegraphics[height=8cm]{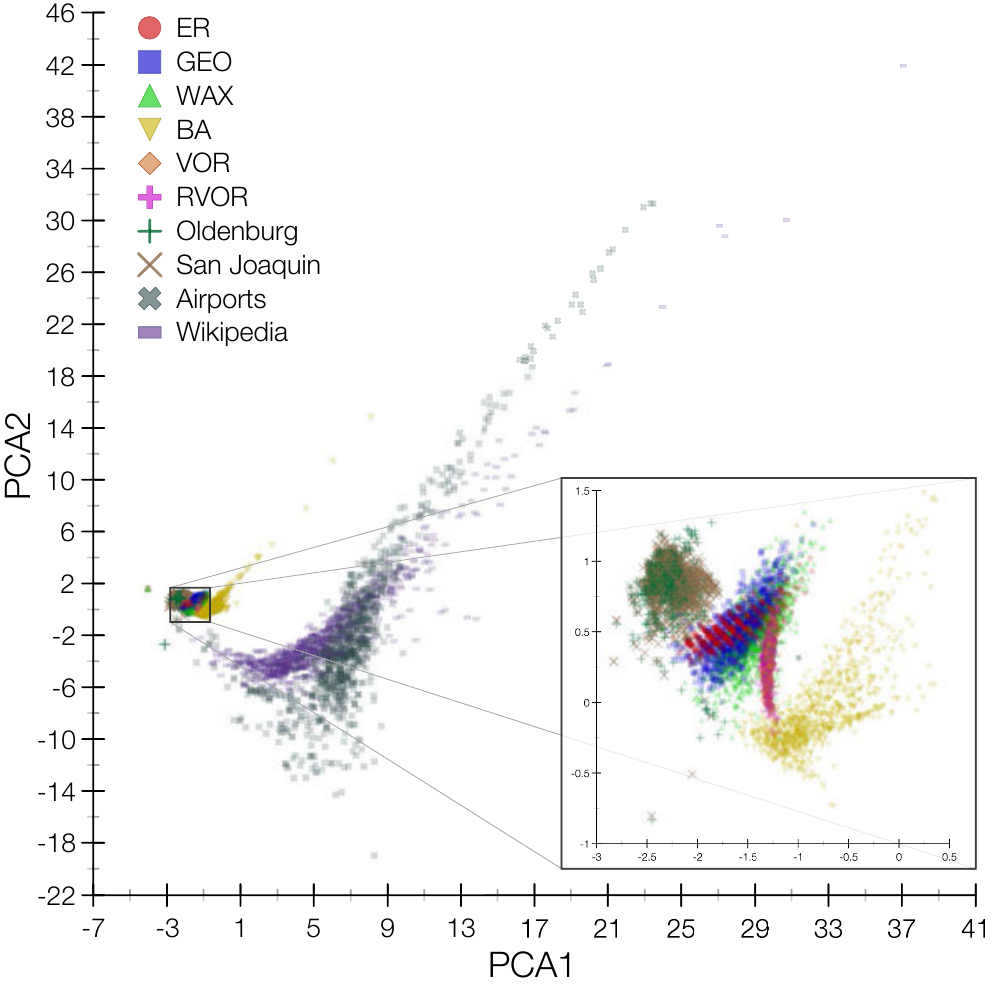} }
\end{center}
\caption{(Color online) PCAs obtained from the standardized backbone and merged symmetries from level $h=2$ to $h=4$ (a), and from concentric measurements \cite{Costa:2006p278} considering the same range of levels (b). Because the networks present distinct number of nodes, only $1000$ nodes randomly sampled from each network are considered for the calculation.}
\label{Fig:PCA_onlySymmetry}
\end{figure*}

To obtain a concise comparison of the symmetry results obtained for all networks studied here, we apply the Principal Component Analysis \cite{jolliffe:2002} (PCA). Because the networks do not have the same number of nodes, we randomly selected 1000 nodes from each network and calculated the backbone and merged symmetries for $h=2,3$ and $4$, defining a 6-dimensional space containing all the selected nodes. The PCA is applied to this data and the first two components are retained, the result is shown in Fig.~\ref{Fig:PCA_onlySymmetry}(a). The PCA in Fig.~\ref{Fig:PCA_onlySymmetry}(a) encompasses most of the main results of this paper and highlights the relationships between the symmetric features of the considered networks. The two first PCA axes have been verified to account for $\approx 86\%$ of the total variance. The component \emph{PCA1} corresponds to a balanced linear combination of all symmetry measurements, i.e.\ $\text{\emph{PCA1}} = 0.37 Sb_i^{(2)} + 0.43 Sm_i^{(2)} +0.44 Sb_i^{(3)}  +0.43 Sm_i^{(3)}  +0.43 Sb_i^{(4)}  +0.34 Sm_i^{(4)}$, as a consequence higher \emph{PCA1} reflects in higher symmetry. The Wikipedia, airports and BA networks present small \emph{PCA1} values (i.e.\ low symmetry), possibly reflecting their power-law nature. GEO and WAX models present little overlap between themselves even though they share a geographical nature. This might be caused by the existence of long-range connections present in the WAX model, which are not allowed in the GEO networks. Both street networks present great overlap in the PCA space and, along \emph{PCA1}, they are closer to the ER model than to the geographical models. At the same time, the RVOR network presented an even larger overlap with the two cities. 

In order to compare the results obtained using the symmetry measurements, we also applied the PCA while taking into account several more traditional concentric characteristics \cite{Costa:2006p278} (e.g. concentric degree, clustering). The result is shown in Fig.~\ref{Fig:PCA_onlySymmetry}(b). By comparing the two PCA projections, it becomes clear that the symmetry measurements provide a more discriminative representation of the several considered networks, reflected in less overlap between clusters allied to wider distributions of nodes within the clusters associated to each network. The relatively small overlaps between some of the networks in Fig.~\ref{Fig:PCA_onlySymmetry}(a), such as between ER and the street networks, are indeed expected. A good deal of such remarkable properties can be understood to be related to the insensitivity of the symmetry measurements to the node degree, which otherwise implies in widely different sizes of clusters (see Fig.~\ref{Fig:PCA_onlySymmetry}(b)).

The spatial dispersion of the node symmetry measurements, quantified in terms of the trace of the respective covariance matrix~\cite{fukunaga1990introduction}, obtained for the several networks considered here, are given in Table \ref{table:spreading} respectively to the concentric levels $h=2, 3,$ and 4.

\begin{table*}[htdp]
\caption{Trace of the covariance matrix calculated for the backbone and merged symmetry values at each concentric level.}
\begin{center}
\begin{tabular}{c c c c c c c c c c c}
    \textbf{\begin{tabular}[c]{@{}c@{}}Concentric\\ Level (h)\end{tabular}}         & \textbf{ER} & \textbf{GEO} & \textbf{WAX} & \textbf{Voronoi} & \textbf{\begin{tabular}[c]{@{}c@{}}Voronoi\\ (Rewired)\end{tabular}} & \textbf{BA}    & \textbf{Oldenburg} & \textbf{San Joaquin} & \textbf{Airports} & \textbf{Wikipedia}  \\ \hline\hline
\textbf{2} & 0.010       & 0.020        & 0.044        & 0.005            & 0.039                                                                & \textbf{0.112} & 0.023              & 0.021                & \textbf{0.182}    & \textbf{0.144}     \\
\textbf{3} & 0.011       & 0.052        & 0.034        & 0.013            & 0.069                                                                & 0.021          & 0.038              & 0.038                & 0.081             & 0.033              \\
\textbf{4} & 0.067       & 0.014        & 0.022        & 0.019            & 0.050                                                                & 0.005          & 0.039              & 0.039                & 0.040             & 0.007             \\ \hline
\textbf{All} & 0.088	  & 0.086	& 0.100	& 0.037	& 0.158	& 0.138	& 0.100	& 0.098	& 0.303	& 0.184             \\ \hline
\end{tabular}
\end{center}
\label{table:spreading}
\end{table*}%

Another interesting result comes from the fact that the \emph{PCA1} axis retains $71\%$ of their variance, which allow us to rank the networks in terms of an overall symmetry value, i.e.\ \emph{PCA1}. The most symmetric networks are those of streets, followed by the ER model and the two spatial models. On the other hand, Wikipedia, airports and BA models are the more asymmetrical networks and also present the higher dispersion values for $h=2$ (please see Table \ref{table:spreading}). Such ordering along \emph{PCA1} tends to reflect the regularity/simplicity of the original networks, a possibility that needs to be further investigated.

\section{Conclusions} 

Complex networks can be studied at different topological scales: local-scale characterized by the connectivity between neighbors of a node, intermediate-scale usually associated to communities; and global-scale concerning the overall structure of the network. Several methods have been developed for the characterization of global-scale symmetries in networks, mostly based on automorphism~\cite{godsil2001algebraic,biggs1993algebraic}, which implies some difficulties, specially regarding normalization between different sizes of networks. In addition, developing concepts and methods capable of quantifying the symmetry locally (e.g.\ around a given neighborhood of a node) is of great interest because they reflect the topological structure most closely affecting the dynamical and topological properties at each node in the network.

The concept of symmetry developed in the current article is associated to the regularity of the transition probabilities of a concentric random walk. Since the calculations are always done using a single node as a reference, we can relate the concept of topological symmetry with our intuitive visual perception of radial symmetry. This relationship is particularly noticeable for the Erd\H{o}s-Renyi model (ER). Since connections inside the same concentric levels are uncommon for the ER network, which give rise to a tree-like organization, the topological symmetry is mainly determined by the balance of connections along tree branches. This property also contributes for a clear visualization of the topological symmetry. 

The Barab\'asi Albert (BA) model provides a contrast to the ER model, since it is highly probable that one or more hubs are accessed after one or two steps from the reference node. This property implies a strong unbalance on the transition probabilities of the concentric random walk, since nodes reached after passing through the hub are accessed with much lower probability than other nodes on the same concentric level that are not associated with hubs. Another interesting network model investigated here are spatial networks. Since the nodes in these networks have a well-defined position influencing the connectivity of patterns, it is interesting to quantify how this influence translates into topological symmetry.

Regarding real-world networks, the airport and Wikipedia showed a wider range of symmetry values than any network model. This might happen because the connectivity patterns of the nodes do not follow single rules as in the models, in the sense that they present a hybrid topology and nonvanishing degree-degree correlations. For the airport networks, different countries or states have specific demands (e.g. passenger flow). In the Wikipedia, each subject (i.e., mathematics, biology, etc) tends to present specific citation patterns~\cite{Silva2011431}. Both the ER and street networks have highly symmetric patterns, even though they differ structurally, since street networks present high occurrence of cycles while ER have locally tree-like patterns. Still, the topological patterns are more symmetric than those observed in the models, which is confirmed by higher values of backbone and merged symmetry. This is probably a consequence of the adjacency constraints implied by planarity that needs to be followed in city planning. This effect is even more accentuated on Voronoi network~\cite{barthelemy2011spatial}. Although the GEO network share many similar characteristics with the street networks, they are not specifically made to be navigable.

The PCA provided an important overview of the symmetry measurements, revealing that the described framework can enrich the characterization of complex networks at the local scale as well as on a system-wide perspective. In conjunction with the correlation analysis we confirmed that the new measurements unveiled new information since they presented low correlation with traditional features. Furthermore, the two symmetry indices (backbone and merged) were found to be complementary, by being little correlated. For instance, the backbone and merged symmetries obtained from Wikipedia and Airport networks were widely distributed in the two-dimensional space defined by the two symmetry indices. The same conclusion holds for the symmetry considering distinct concentric levels, as became clear by looking at the first principal component of the PCA.

As developed in this work, the symmetry measurements were found to capture many distinct concentric patterns of connectivity. A natural extension of our analysis is to verify which of these patterns occur more often than it would be expected by chance on real-world networks. Therefore, by analyzing the density of points in the symmetry space we can identify statistical motifs in the network structure. It would be important to verify how many distinct patterns can be associated to the same pair of symmetry values. Another additional development is to identify typical symmetry patterns specific to distinct regions of the real-world networks. For example, would a given region of the airport network present some intrinsic organization? Or, would given areas of Wikipedia exhibit typical symmetry patterns?

The symmetry of a network can also influence many types of dynamics, such as how a disease propagate over the network. Diffusion and disease dynamics are of particular interest since it is possible to analyze how the number of symmetric patterns in the network impacts the speed of a dynamics evolution.  It would be interesting to extend previous investigations showing that degree heterogeneity can facilitate disease propagation on networks~\cite{PhysRevLett.86.3200,PhysRevLett.109.098102}.

\section*{Acknowledgements}

F. N. Silva acknowledges CAPES. C. H. Comin thanks FAPESP (Grant No. 11/22639-8) for financial support. T. K. D. M. Peron acknowledges FAPESP (Grant no. 2012/22160-7) for support. F. A. Rodrigues acknowledges CNPq (grant 305940/2010-4), FAPESP (grant 2011/50761-2 and 2013/26416-9) and NAP eScience - PRP - USP for financial support. L. da F. Costa thanks CNPq (Grant no. 307333/2013-2) and NAP-PRP-USP for support. This work has been supported also by FAPESP grants 12/50986-7 and 11/50761-2.

\bibliographystyle{apsrev}
\bibliography{paper}

\end{document}